\documentclass[manuscript,screen,acmsmall,nonacm]{acmart}

\AtBeginDocument{%
  \providecommand\BibTeX{{%
    \normalfont B\kern-0.5em{\scshape i\kern-0.25em b}\kern-0.8em\TeX}}}

\setcopyright{acmcopyright}
\copyrightyear{2022}
\acmYear{2022}
\acmDOI{TBD}

\usepackage{xspace}
\usepackage{ifthen}
\newcommand\LTLg{\mathcal{G}}
\newcommand\LTLf{\mathcal{F}}
\usepackage{xcolor}
\usepackage{colortbl}
\usepackage{cleveref}

\usepackage[framemethod=TikZ]{mdframed}

\usepackage{graphbox}
\usepackage{mathtools}
\usepackage[most]{tcolorbox}
\usepackage{fancybox}
\usepackage{makecell}
\usepackage{multirow}

\newboolean{showcomments}
\setboolean{showcomments}{true}
\ifthenelse{\boolean{showcomments}}
{\newcommand{\nb}[2]{
  \fcolorbox{black}{yellow}{\bfseries\sffamily\scriptsize#1}
  {\sf\small$\blacktriangleright$\textit{#2}$\blacktriangleleft$}
 }
 
}
{\newcommand{\nb}[2]{}
 
}

\usepackage{subcaption}
\usepackage{epstopdf}

\newtcolorbox{mybox}{colback=red!5!white,colframe=red!75!black}
\newcommand{\simulink}{Simulink\textsuperscript{\tiny\textregistered}\xspace}

\usepackage[binary-units,per-mode=symbol,detect-weight=true, detect-family=true,inter-unit-product =\cdot]{siunitx}

\newcommand\timedomain{\ensuremath{\mathbb{T}}}
\newcommand\real{\ensuremath{\mathbb{R}}\xspace}

\newcommand{\Breach}{\textsc{Breach}\xspace}
\newcommand{\FalStar}{\textsc{FalStar}\xspace}

\acmBooktitle{Transactions on Software Engineering and Methodology}
\acmPrice{}
\acmISBN{}

\usepackage[inline]{enumitem}

\definecolor{myBlue}{RGB}{0,133,255}
\newcommand\phase[1]{\tikz[baseline=(X.base)]\node [draw=black,fill=white,thick,rectangle,inner sep=2pt, rounded corners=2pt](X){\color{black}\textbf{#1}};}

\newcommand{\system}{\ensuremath{S}\xspace}
\newcommand{\fitness}{\ensuremath{f}\xspace}

\newcommand{\asmpt}{\ensuremath{A}\xspace}

\newcommand{\budget}{\ensuremath{T}\xspace}

\newcommand{\requirement}{\ensuremath{\varphi}}

\newcommand\inputs{\ensuremath{\texttt{i}}\xspace}
\newcommand\inputsignal{\ensuremath{i}}
\newcommand\outputs{\ensuremath{\texttt{o}}\xspace}
\newcommand\outputsignal{\ensuremath{o}}

\newcommand{\NAME}{\texttt{ATheNA}\xspace}
\newcommand{\NAMEMANUAL}{\texttt{ATheNA-SM}\xspace}

\newcommand{\NAMEAUTOMATIC}{\texttt{ATheNA-SA}\xspace}

\newcommand{\NAMESIMULINK}{\texttt{ATheNA-S}\xspace}

\newcommand{\FalStarTOOL}{\textsc{FalStar}\xspace}
\newcommand{\foresee}{\textsc{ForeSee}\xspace}

\newcommand{\falsify}{falsify\xspace}

\newcommand{\FalCAuN}{\texttt{FalCAuN}\xspace}
\newcommand{\staliro}{\texttt{S-Taliro}\xspace}
\newcommand{\Aristeo}{\texttt{Aristeo}\xspace}

\newcommand{\nruns}{\ensuremath{50}\xspace}
\newcommand{\niterations}{\ensuremath{300}\xspace}

\newcommand{\RQonenumathenawins}{\ensuremath{29}\xspace}

\newcommand{\RQonewinsathenavsstaliro}{\ensuremath{7.7\%}\xspace}
\newcommand{\RQonewinsathenavsstaliromax}{\ensuremath{60\%}\xspace}
\newcommand{\RQonewinsathenavsstaliromin}{\ensuremath{0\%}\xspace}
\newcommand{\RQonewinsathenavsstalirostd}{\ensuremath{12.8\%}\xspace}

\newcommand{\RQonewinsathenavsmanual}{\ensuremath{11.2\%}\xspace}
\newcommand{\RQonewinsathenavsmanualmax}{\ensuremath{58\%}\xspace}
\newcommand{\RQonewinsathenavsmanualmin}{\ensuremath{0\%}\xspace}
\newcommand{\RQonewinsathenavsmanualstd}{\ensuremath{18.0\%}\xspace}

\newcommand{\RQonenumathenaloses}{\ensuremath{10}\xspace}
\newcommand{\RQonelosesathenavsstaliro}{\ensuremath{4.0\%}\xspace}
\newcommand{\RQonelosesathenavsstaliromax}{\ensuremath{6\%}\xspace}
\newcommand{\RQonelosesathenavsstaliromin}{\ensuremath{2\%}\xspace}
\newcommand{\RQonelosesathenavsstalirostd}{\ensuremath{2.0\%}\xspace}

\newcommand{\RQonelosesathenavsmanual}{\ensuremath{10.6\%}\xspace}
\newcommand{\RQonelosesathenavsmanualmax}{\ensuremath{24\%}\xspace}
\newcommand{\RQonelosesathenavsmanualmin}{\ensuremath{2\%}\xspace}
\newcommand{\RQonelosesathenavsmanualstd}{\ensuremath{7.8\%}\xspace}

\newcommand{\RQoneathenavsstaliro}{\ensuremath{7.9\%}\xspace}
\newcommand{\RQoneathenavsstaliromax}{\ensuremath{60\%}\xspace}
\newcommand{\RQoneathenavsstaliromin}{\ensuremath{-6\%}\xspace}
\newcommand{\RQoneathenavsstalirostd}{\ensuremath{13.1\%}\xspace}

\newcommand{\RQoneathenavsmanual}{\ensuremath{8.1\%}\xspace}
\newcommand{\RQoneathenavsmanualmax}{\ensuremath{58\%}\xspace}
\newcommand{\RQoneathenavsmanualmin}{\ensuremath{-24\%}\xspace}
\newcommand{\RQoneathenavsmanualstd}{\ensuremath{19.5\%}\xspace}

\newcommand{\minvariation}{\ensuremath{2\%}\xspace}
\newcommand{\maxvariation}{\ensuremath{62\%}\xspace}
\newcommand{\avgvariation}{\ensuremath{23.6\%}\xspace}
\newcommand{\stdvariation}{\ensuremath{19.2\%}\xspace}

\newcommand{\RQonenumberofDays}{\ensuremath{35}\xspace}

\newcommand{\numrequirements}{\ensuremath{27}\xspace}
\newcommand{\numrequirementassumption}{\ensuremath{39}\xspace}

\newcommand{\RQonewinsathenabestvsstaliro}{\ensuremath{11.9\%}\xspace}
\newcommand{\RQonewinsathenabestvsstaliromin}{\ensuremath{0\%}\xspace}
\newcommand{\RQonewinsathenabestvsstaliromax}{\ensuremath{60\%}\xspace}
\newcommand{\RQonewinsathenabestvsstalirostd}{\ensuremath{14.4\%}\xspace}

\newcommand{\RQonewinsathenabestvsmanual}{\ensuremath{12.1\%}\xspace}
\newcommand{\RQonewinsathenabestvsmanualmin}{\ensuremath{0\%}\xspace}
\newcommand{\RQonewinsathenabestvsmanualmax}{\ensuremath{62\%}\xspace}
\newcommand{\RQonewinsathenabestvsmanualstd}{\ensuremath{19.0\%}\xspace}

\newenvironment{Answer}[1][]{%
  \ifstrempty{#1}%
  {\mdfsetup{%
    frametitle={%
      \tikz[baseline=(current bounding box.east),outer sep=0pt]
      \node[line width=0pt,anchor=east,rectangle,draw=white,fill=white]
    ;}}
  }%
  {\mdfsetup{%
    frametitle={%
      \tikz[baseline=(current bounding box.east),outer sep=0pt]
      \node[anchor=east,rectangle,draw=white,fill=white]
    {\strut #1};}}%
  }%
  \mdfsetup{innertopmargin=-5pt,linecolor=black,%
            linewidth=0.5pt,topline=true,%
            frametitleaboveskip=\dimexpr-\ht\strutbox\relax,}
  \begin{mdframed}[]\relax%
  }{\end{mdframed}}

\begin{document}


\title{Search-based Software Testing Driven by Automatically Generated and Manually Defined Fitness Functions
}

\author{Federico Formica}
\email{formicaf@mcmaster.ca}
\affiliation{
  \institution{McMaster University}
  \city{Hamilton}
  \country{Canada}
}

\author{Tony Fan}
\email{fant6@mcmaster.ca}
\affiliation{
  \institution{McMaster University}
  \city{Hamilton}
  \country{Canada}
}

\author{Claudio Menghi}
\email{claudio.menghi@unibg.it}
\affiliation{
  \institution{University of Bergamo}
  \city{Bergamo}
  \country{Italy}
}
\affiliation{%
  \institution{McMaster University}
  \city{Hamilton}
  \country{Canada}
}

\renewcommand{\shortauthors}{Federico Formica, Tony Fan, Claudio Menghi}

\begin{abstract}
Search-based software testing (SBST) typically relies on fitness functions to guide the search exploration toward software failures. 
There are two main techniques to define fitness functions: (a)~automated fitness function computation from the specification of the system requirements, and (b)~manual fitness function design.
Both techniques have advantages.
The former uses information from the system requirements to guide the search toward portions of the input domain more likely to contain failures.
The latter uses the engineers' domain knowledge.

We propose \NAME, a novel SBST framework that combines fitness functions automatically generated from requirements specifications and those manually defined by engineers. 
We design and implement \NAMESIMULINK, an instance of \NAME that targets \simulink models. 
We evaluate \NAMESIMULINK by considering a large set of models from different domains. 
Our results show that \NAMESIMULINK generates more failure-revealing test cases than existing baseline tools and that the difference between the runtime performance of \NAMESIMULINK and the baseline tools is not statistically significant.
We also assess whether \NAMESIMULINK could generate failure-revealing test cases when applied to two representative case studies: one from the automotive domain and one from the medical domain. 
Our results show that \NAMESIMULINK successfully revealed a requirement violation in our case studies.
\end{abstract}

\ccsdesc[500]{Software and its engineering}
\ccsdesc[500]{Software and its engineering~Software testing and debugging}

\keywords{Testing, Falsification, Fitness Functions, CPS}

\maketitle

\section{Introduction}
\label{sec:Intro}
Software failures in cyber-physical systems (CPS) can have \emph{catastrophic and costly consequences}. 
For example, automotive software failures led to severe injuries and the loss of human lives (e.g., ~\cite{Uber,GMDefect,TeslaCrash}). Car manufacturers had to recall their vehicles, causing reputation damage and millions of U.S. Dollars lost (e.g.,~\cite{GMDefect,HondaDefects,TeslaDefects,StellantiDefects,LandRover,recalls,costs2}). 
To prevent these scenarios, CPS engineers extensively test their systems to \emph{detect safety-critical software failures}~\cite{10.1145/2631890.2631891,garousi2018testing,zhou2018review,duan2018systematic}. 
Automated testing tools facilitate this activity (e.g., ~\cite{Aristeo,S-Taliro,matinnejad2015search,abdessalem2018testing}). 
These tools are regularly used in safety-critical CPS domains, including automotive (e.g.,~\cite{matinnejad2015search}), aerospace (e.g.,~\cite{Aristeo}), and medical (e.g.,~\cite{majikes2013literature}).

Automated testing often (e.g.,~\cite{Aristeo}) relies on \emph{search-based software testing} (SBST). 
SBST iteratively generates test cases until a software failure is detected or the SBST framework exceeds the time budget allotted for the testing activity. 
It relies on different (a)~optimization algorithms (e.g.,~\cite{Mathesen2019,Cohen2021}), (b)~input types (e.g.,~\cite{Ramezani2021}), (c)~surrogate models (e.g.,~\cite{NEJATI2023107286,Aristeo}),
and (d)~fitness functions (e.g.,~\cite{matinnejad2015search}).
This paper considers the fitness function design, a challenging task for designing effective SBST frameworks (\cite{DBLP:journals/stvr/SalahiradAG20,10.1145/2001858.2001929,6032439,sahin2016comparisons,8819910,DBLP:journals/ase/AletiMG17,9159064,sadri2022survey}).

\emph{Fitness functions} guide SBST frameworks in generating new test cases. They provide metrics (a.k.a. fitness values) that estimate how close the test cases are to detecting a failure~\cite{1357891}. 
To effectively and efficiently generate failure-revealing test cases, it is critical to select appropriate fitness functions~\cite{10.1145/2001858.2001929,6032439,sahin2016comparisons,8819910}.
Fitness landscape analysis activities evaluate how the fitness value changes over the search space and usually support the fitness function design (e.g.~\cite{kim2003new,pohlheim1999visualization, DBLP:journals/ase/AletiMG17}).
Fitness landscape analysis can help understand the search process and its probability of success (e.g.~\cite{hart2001gavel,kim2002visualization,kim2003new}).
However, despite the breadth and diversity of testing domains and solutions, the design of fitness functions is still complex and time-consuming (e.g.,~\cite{DBLP:journals/stvr/SalahiradAG20,9159064,10.1007/978-3-319-09940-8_3}). 

There are two mainstream techniques to define fitness functions:
\emph{automated generation} and \emph{manual definition}.

\emph{Automated generation of the fitness function} (e.g.,~\cite{Fainekos2019,Donze2010,Varnai2020,9155827,Pant2017,Mehdipour2019,Lindemann2019}) derives the fitness function from other artifacts without any human intervention.
For example, many SBST frameworks (e.g.,~\cite{S-Taliro,Aristeo,Falstar,ernst2021falsification})
use fitness functions to compute the robustness values derived from temporal logic specifications expressing system requirements (e.g.,~\cite{Fainekos2019,menghi2019generating,Pant2017,Haghighi2019}).
These functions typically use the requirement structure to guide the search toward portions of the input domain more likely to lead to system failures. 
Automatically generated fitness functions support SBST in the detection of failures (e.g.,~\cite{Donze2010, Varnai2020,Haghighi2019,9155827}).
They were used to detect failures in industrial models (e.g.,~\cite{Aristeo,jin2014powertrain}), and are used within international tool competitions (e.g.,~\cite{DBLP:conf/arch/ErnstABCDFFG0KM21}).
Automatically generated fitness functions are \emph{general purpose} and do not use engineers' domain knowledge for the fitness value computation.

Unlike the automated approach, 
\emph{manual fitness function definition} (e.g., ~\cite{abdessalem2018testing,10.1145/2001858.2001929,7886937,8453180,10.1145/2642937.2642978}) requires engineers to design the fitness functions using their experience and domain knowledge.
For example, engineers can 
manually define fitness functions to guide the search toward the generation of inputs that are more likely to show the violation of liveness, stability, smoothness, and responsiveness requirements~\cite{matinnejad2015search}. 
Manually defined fitness functions effectively and efficiently support SBST --- they enable the detection of failures that domain experts could not find by manual testing (e.g., ~\cite{matinnejad2015search}).
Manual fitness function design enables engineers to write \emph{model-specific} fitness functions that guide the search toward specific areas that are more likely to contain failures, e.g., the boundaries of the input domain.  
However, manually defined fitness functions are biased and, in some cases, may concentrate the search on areas of the input domains that do not contain failures.

This work proposes \NAME (AuTomatic-maNuAl search-based testing), a novel black-box \emph{SBST framework driven by automatically generated and manually defined fitness functions}. 
\NAME combines the benefits of automatically generated and manually defined fitness functions: it exploits both the structure of the requirements and the engineers' domain knowledge to guide the search toward specific areas of the input domain that are likely to reveal software failures.
In addition, we define \NAMESIMULINK, an instance of \NAME that supports \simulink models.
We consider \simulink models since they are widely used for specifying the behavior of CPSs~\cite{boll2021characteristics,DBLP:journals/sosym/LiebelMTLH18} in a variety of domains, including automotive~\cite{matinnejad2015search}, energy~\cite{jin2014powertrain}  and medical~\cite{sankaranarayanan2012simulating}.
We implement \NAMESIMULINK as a plugin for S-Taliro~\cite{S-Taliro}, a well-known SBST framework for \simulink models recently classified as ready for industrial usage~\cite{kapinski2016simulation}.
\NAME is the first solution that enables engineers to combine manual and automatic fitness function design.
Enabling the combination of the two fitness functions is a significant contribution to the software engineering field since it allows engineers to exploit the benefits of the combined fitness function that, as we show in our evaluation, outperforms existing solutions.

We evaluate the effectiveness and efficiency of \NAMESIMULINK in generating failure-revealing test cases.
We compare \NAMESIMULINK with \staliro~\cite{S-Taliro}, a tool
that supports automatically generated fitness functions,
and \NAMEMANUAL, a customization of \NAMESIMULINK that supports manually defined fitness functions.
We consider seven models and $27$ requirements from ARCH~2021~\cite{DBLP:conf/arch/ErnstABCDFFG0KM21,ARCHWEBSITE}, an international SBST competition for \simulink models held as a part of the international conference on Computer Safety, Reliability, and Security (SAFECOMP)~\cite{DBLP:conf/safecomp/2021}.
For each requirement, we consider a set of assumptions for the inputs of the model.
In total, we compare the tools by considering $39$ assumption-requirement combinations.
We considered two versions of \NAME that use different
functions to combine the manual and automatic fitness values.
Our results show that (a)~\NAMESIMULINK performs better than the baseline tools for most of our assumption-requirement combinations ($\approx74\%$ and $100\%$ for the two versions of \NAME we considered), (b)~\NAMESIMULINK  generates more failure-revealing test cases than \staliro (+\RQoneathenavsstaliro and 
+\RQonewinsathenabestvsstaliro) and \NAMEMANUAL (+\RQoneathenavsmanual and +\RQonewinsathenabestvsmanual 
), and (c)~the difference between the runtime performance of \NAMESIMULINK and the baseline tools is not statistically significant for one version of \NAME and negligible for the other.
Additionally, we assess how applicable and useful is \NAMESIMULINK in generating failure-revealing test cases for two large  \simulink models.
The first model is an electrical automotive software control system~\cite{automotive} developed as a part of the EcoCAR Mobility Challenge~\cite{EcoCAR}, a competition sponsored by the U.S. Department of Energy~\cite{DOE}, General Motors~\cite{GM}, and MathWorks~\cite{MathWorks}.
The second model is mechanical ventilator~\cite{Ventilator} developed by MathWorks~\cite{MathWorks}.
We evaluate whether \NAMESIMULINK could generate any failure-revealing test cases. 
Our results show that \NAMESIMULINK could generate a failure-revealing test within practical time limits for both the automotive and the ventilator models.
We present and discuss the problems identified by the failure-revealing test cases returned by \NAMESIMULINK.

To summarize, our contributions  are:
\begin{itemize}[leftmargin=*]
    \item We propose the \NAME framework (Section~\ref{sec:Athena});
    \item We define \NAMESIMULINK, an instance of \NAME that supports \simulink models (Section~\ref{sec:applSimulink});
    \item We implement \NAMESIMULINK  as a plugin for \staliro (Section~\ref{sec:implementation});
    \item We empirically assess the benefits of  \NAME  (Section~\ref{sec:evaluation});
    \item We discuss the impact of our findings on the software engineering practice (Section~\ref{sec:discussion}).
\end{itemize}

This work is organized as follows: 
Section~\ref{sec:Athena} presents \NAME.
Section~\ref{sec:applSimulink} describes the instance of \NAME that targets \simulink models.
Section~\ref{sec:implementation} provides implementation details.
Section~\ref{sec:evaluation} empirically assesses our contribution.
Section~\ref{sec:discussion} discusses our findings and presents threats to validity.
Section~\ref{sec:RelWorks} presents related work.
Section~\ref{sec:conclusion} concludes the work.


\section{Automatic-Manual SBST}
\label{sec:Athena}
\Cref{fig:bee} provides an overview of the \NAME (AuTomatic-maNuAl) search-based testing framework.
Squared boxes report the steps of \NAME.
Incoming and outgoing arrows describe the inputs and outputs of the different steps. Arrows with no source represent the inputs of the \NAME framework.
Arrows with no destination represent the outputs of the \NAME framework.
Arrows connecting two boxes link subsequent steps.

\tikzstyle{output} = [coordinate]

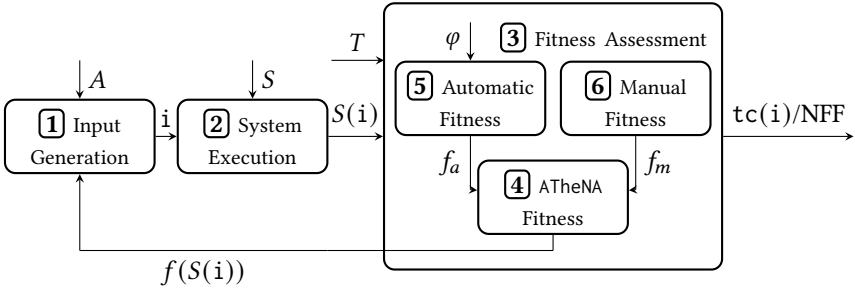
\begin{figure}[t]
\begin{tikzpicture}[auto,
 block/.style ={rectangle, draw=black, thick, fill=white!20, text width=5em,align=center, rounded corners},
 block1/.style ={rectangle, draw=blue, thick, fill=blue!20, text width=5em,align=center, rounded corners, minimum height=2em},
 line/.style ={draw, thick, -latex',shorten >=2pt},
 cloud/.style ={draw=red, thick, ellipse,fill=red!20,
 minimum height=1em}]
 
\draw (0,0) node[block] (Input) {\phase{1} \footnotesize Input \\ Generation};
\node[block, right of=Input,node distance=2.3cm] (Model){\phase{2} \footnotesize System\\  Execution};
 \node [block, right of=Model,node distance=4cm, minimum height=3.5cm,minimum width=4.5cm] (OverallFitness) {};
 \node [block,above of=OverallFitness,node distance=1.3cm, line width=0mm,draw=white!80,text width=4.1cm] (OverallFitnessLabel) {\phantom{xxxxxxx} \phase{3} \footnotesize Fitness Assessment};
 \node [output, above of=OverallFitness,node distance=0.5cm] (ftn) {};
\node[block, below of=ftn,node distance=1.3cm] (Requirement){\phase{4} \footnotesize \NAME\\ Fitness
};
 \node [block, right of=ftn,node distance=1.1cm] (ManualFitness) {\phase{6} \footnotesize Manual\\  Fitness};
\node [block, left of=ftn,node distance=1.1cm] (AutomaticFitness) {\phase{5} \footnotesize Automatic\\ Fitness};

\node [output, right of=OverallFitness,node distance=4cm] (uoutput) {};
\node [output, above of=Input,node distance=1cm] (Constraints) {};
 \node [output, above of=Model,node distance=1cm] (InputModel) {};

\node [output, below of=Requirement,node distance=0.7cm] (v1a) {};

\node [output, left of=v1a,node distance=3cm] (v1) {};
 
\node [output, left of=v1,node distance=3.3cm] (v2) {};
 
 \node [output, below of=ManualFitness,node distance=1.2cm] (mftwo) {};

\node [output, below of=AutomaticFitness,node distance=1.2cm] (mf) {};

\node [output, above of=AutomaticFitness,node distance=1cm] (ff) {};

\node [output, left of=OverallFitness,node distance=2.25cm] (budget1) {};

\node [output, above of=budget1,node distance=1cm] (budget2) {};

\node [output, left of=budget2,node distance=0.7cm] (budget3) {};

 \draw[-stealth] (budget3) -- (budget2)
    node[midway,above]{$\budget$};

\draw[-stealth] (Input.east) -- (Model.west)
    node[midway,above]{$\inputs$};
\draw[-stealth] (Constraints.south) -- (Input.north)
    node[midway,right]{$\asmpt$};
\draw[-stealth] (InputModel.south) -- (Model.north)
    node[midway,right]{$\system$};

\draw[-] (v1) -- (v2)node[midway,below]{$\fitness(\system(\inputs))$};
\draw[-stealth] (v2) -- (Input.south);
 \draw[-stealth] (Model) -- (OverallFitness.west) node[midway,above]{$\system(\inputs)$};
 \draw[-stealth] (OverallFitness.east) -- (uoutput) node[midway,above]{$\texttt{tc}(\inputs)$/NFF};
  \draw[-] (AutomaticFitness.south) -- (mf) node[midway,left]{$\fitness_a$};
  \draw[-stealth] (mf) -- (Requirement) node[midway,left]{};
\draw[-] (ManualFitness.south) -- (mftwo) node[midway,right]{$\fitness_m$};
  \draw[-stealth] (mftwo) -- (Requirement) node[midway,left]{};
 \draw[-stealth] (ff) -- (AutomaticFitness) node[midway,left]{$\requirement$};
 
   \draw[-] (Requirement) -- (v1a) node[midway,left]{};
 
  \draw[-] (v1a) -- (v1) node[midway,left]{};
 \end{tikzpicture}
\caption{Overview of the \NAME testing framework.}
\label{fig:bee}
\Description[Flowchart representing the steps of the \NAME testing framework.]{Flowchart representing the connection between the six steps of the \NAME testing framework. The main iteration loop is formed by Steps 1, 2, and 3. Step 3 is further split into Steps 4, 5, and 6. Steps 5 and 6 are executed in parallel and feed into Step 4.}
\end{figure}

\NAME has four inputs: 
a model of the system to be tested~(\system); 
an assumption on the system inputs~($\asmpt$), 
a time budget~($\budget$), and a requirement ($\requirement$).
The output of \NAME is a failure-revealing test case ($\texttt{tc}(\inputs)$) or an indication that no failure-revealing test case was found (NFF --- No Failure Found) within the time budget.

To detect a failure-revealing test case, \NAME iteratively repeats the steps presented in~\Cref{fig:bee}:
\begin{itemize}[leftmargin=*]
    \item \emph{Input Generation} (\phase{1}). It generates an input~($\inputs$) for the system model~($\system$) that satisfies the assumption~($\asmpt$);
    \item \emph{System Execution} (\phase{2}). It runs the system model ($\system$) with the generated input ($\inputs$) and obtains a system execution ($\system(\inputs)$).
    \item \emph{Fitness Assessment} (\phase{3}). It computes the fitness value ($\fitness(\system(\inputs))$) associated with the obtained system execution ($\system(\inputs)$) and assesses whether the fitness value is below a threshold value. 
\end{itemize}
A test case ($\texttt{tc}(\inputs)$) associated with the input ($\inputs$) is failure-revealing if:  (a)~the input satisfies the assumption ($\asmpt$), i.e., $\inputs \models \asmpt$, and (b)~the fitness value ($\fitness(\system(\inputs))$) is smaller than a threshold value (typically the value $0$).
Typically, the fitness value is negative if the property is violated and positive otherwise.
In addition, the higher the positive value computed by $f_a$, the further the system is from violating its requirement, while lower negative values indicate that the system is further from satisfying its requirement~\cite{fainekos2006robustness,Fainekos2019,Falstar,menghi2019generating,DBLP:conf/arch/ErnstABCDFFG0KM21}.

The \NAME framework terminates when a failure-revealing test case ($\texttt{tc}(\inputs)$) is detected or when the framework exceeds the time budget ($\budget$) without finding any failure-revealing test case.
For the former case, \NAME returns the failure-revealing test case. For the latter case, \NAME returns the NFF value.

The fitness value ($\fitness(\system(\inputs))$) is used to guide the \NAME framework. 
The search algorithm tries to find an input associated with a negative fitness value. 
To reach this goal, it uses the fitness value computed in step~\phase{3} to drive the generation of the next input~(\phase{1}).\\
The \emph{Input Generation}, \emph{System Execution}, and \emph{Fitness Assessment} components are shared by many existing SBST tools, such as
\staliro~\cite{S-Taliro},
\Breach~\cite{Breach},
\FalStarTOOL~\cite{ernst2019fast}, 
\FalCAuN~\cite{Falstar},
\falsify~\cite{yamagata2020falsification},
\FalStar~\cite{ernst2019fast},
and \foresee~\cite{falsQBRobCAV2021}.

To compute the fitness value, \NAME combines manually defined and automatically generated fitness functions and has the following steps:
\begin{itemize}[leftmargin=*]
    \item \NAME \emph{Fitness} (\phase{4}). It returns a fitness function ($f$) that combines the values computed by the manually defined and automatically generated fitness functions. Depending on the testing necessities, the fitness function $f$ can prioritize one of the two values. The function can also change the prioritization policy dynamically during the search if the fitness value is not effectively guiding the SBST framework.
     \item \emph{Automatic Fitness} (\phase{5}). It returns a fitness function ($f_a$) automatically generated from requirement~$\requirement$. The requirement is automatically compiled into a function that, given a system execution $\system(\inputs)$, computes a fitness value~($\fitness_a$).
    \item \emph{Manual Fitness} (\phase{6}). It returns a fitness function ($f_m$) manually defined by the engineers. It computes a fitness value for  the system execution $\system(\inputs)$.
\end{itemize}

\NAME can be instantiated by considering different modeling formalisms. 
For \NAME to be applicable for a CPS, that CPS is assumed to be a reactive system, i.e., it takes some inputs (produced by the \emph{Input Generation} component) and produces some outputs that can be monitored (used by the \emph{Fitness Assessment} component).
Alternative instances of \NAME differ in the implementation of the 
\emph{Input Generation}, 
\emph{System Execution},
\emph{Fitness Assessment},
\emph{\NAME Fitness},
\emph{Automatic Fitness},
and
\emph{Manual Fitness} components.

One instance of \NAME that targets \simulink models and assumes that the inputs and the outputs of the CPS are signals over time is presented in the next section.



\section{Automatic-Manual SBST for \simulink}
\label{sec:applSimulink}
This section describes \NAMESIMULINK, an instance of \NAME that supports \simulink, a graphical language for model design.

\Cref{fig:ATbenchmark} presents our running example: the variation of the Automatic Transmission~(AT) model provided by Mathworks~\cite{ATBenchmark}, used in the applied verification for continuous and hybrid systems competition~\cite{DBLP:conf/arch/ErnstABCDFFG0KM21,ARCHWEBSITE}.

\simulink provides visual constructs to design a system model~(\system).
\emph{Blocks} typically represent operations and constant values. They are aggregated into subsystems labeled with \emph{ports} that identify the inputs and outputs of the subsystems.
For example, \texttt{Engine} is one of the subsystems of the \simulink model of \Cref{fig:ATbenchmark}. It has two input ports (\texttt{Ti} and \texttt{Throttle}) and one output port (\texttt{Ne}).
\emph{Connections} link input and output ports. 
For example, a connection links the output port \texttt{Ne} of the \texttt{Engine} subsystem to the input port \texttt{Ne} of the \texttt{Transmission} subsystem.  
The inputs and outputs of a \simulink model are represented by the \emph{inports} and \emph{outports} blocks.
For example, in \Cref{fig:ATbenchmark}, there are two inputs (\texttt{Throttle}, and \texttt{Brake}) and three outputs (\texttt{Speed}, \texttt{RPM}, and \texttt{Gear}). 

We instantiated \NAME to support \simulink models as follows:

\begin{figure}[t]
    \centering
    \includegraphics[width = 0.6 \columnwidth]{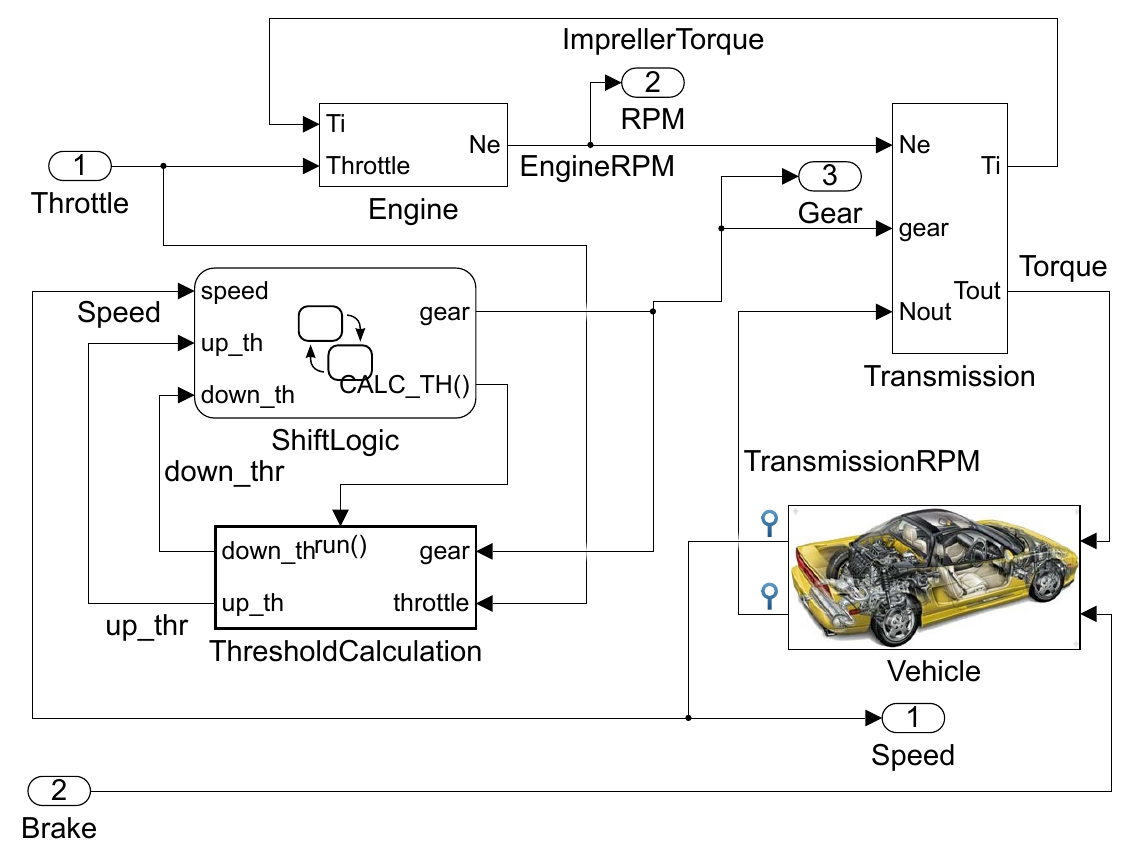}
    \caption{\simulink model of the AT example.}
    \label{fig:ATbenchmark}
    \Description[\simulink model of the AT example.]{\simulink model of the AT example. The model takes as input a Throttle and Brake signals. It returns as output a Speed, Engine RPM, and Gear signals. The model is formed by five subsystems: Engine, Transmission, Vehicle, Shift Logic and Threshold Calculation.}
\end{figure}

\begin{figure*}
    \centering
    \begin{subfigure}[b]{0.49\columnwidth}
        \centering
        \includegraphics[width = \textwidth]{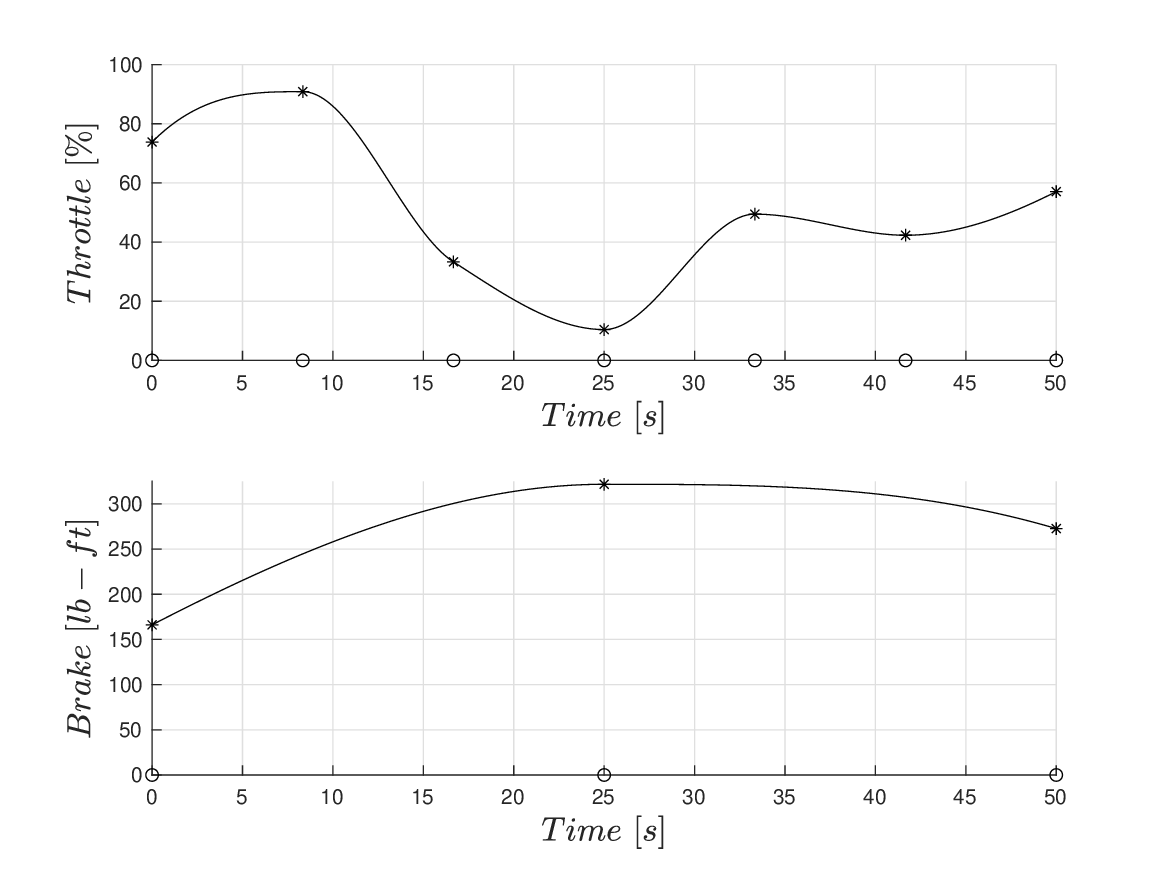}
        \caption{Example of input signals for the AT model between $0s$ and $50s$.}
        \label{fig:InputAT}
        \Description[Example of Throttle and Brake signals for the AT model.]{Example of Throttle and Brake continuous signals for the AT model. The Throttle signal uses seven control points. The Brake signal uses three control points.}
    \end{subfigure}
    \hfill
    \begin{subfigure}[b]{0.49\columnwidth}
        \centering
        \includegraphics[width = \textwidth]{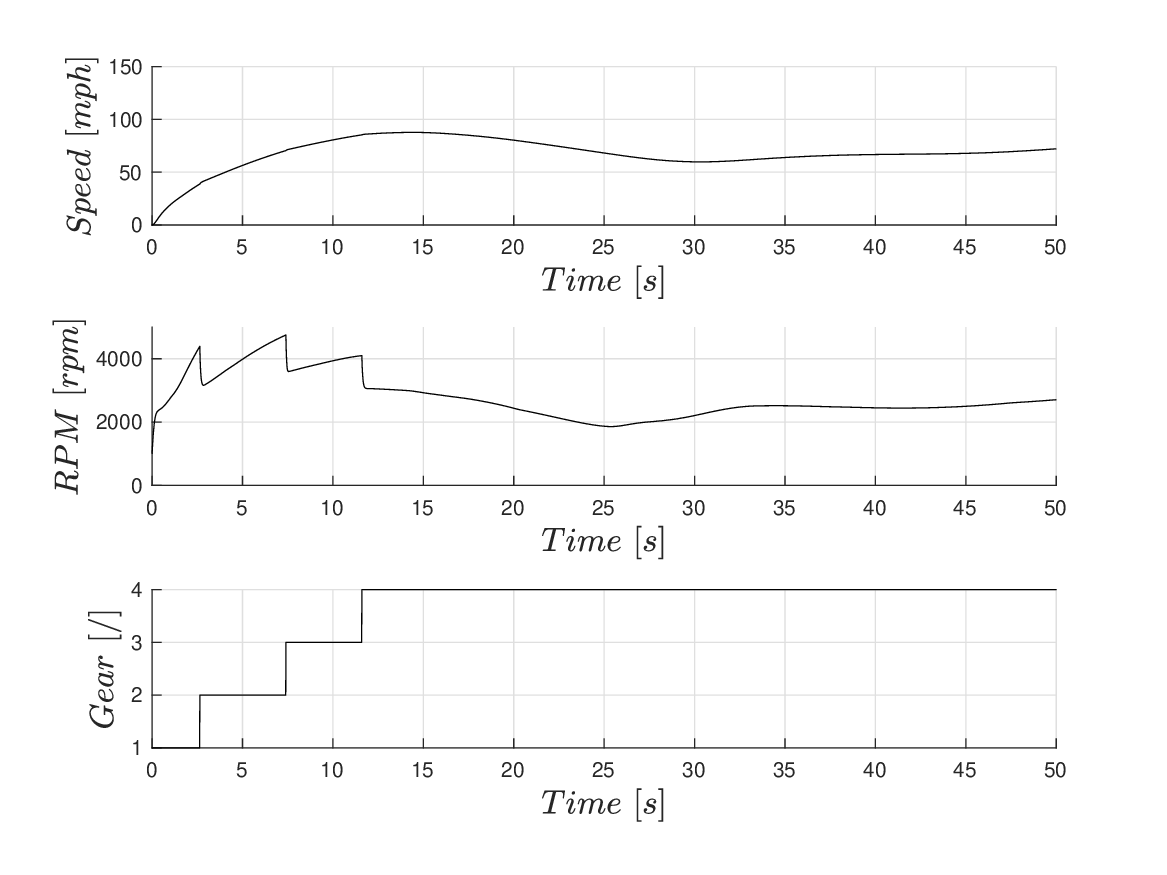}
        \caption{Output signals for the AT  model for the input of \Cref{fig:InputAT}.}
        \label{fig:OutputAT}
        \Description[Example of Speed, Engine RPM, and Gear signals for the AT model.]{Example of Speed, Engine RPM, and Gear signals for the AT model. The plots show the car accelerating until it reaches Gear four and then keeping a constant speed around $70mph$.}
    \end{subfigure}
    
    \caption{Example input and output for the AT \simulink model.}
    \label{fig:InputOutputSignals}
\end{figure*}

\emph{Input Generation} (\phase{1}). 
The input generation step generates a set of signals $\inputs=\{\inputsignal_1, \inputsignal_2, \ldots, \inputsignal_m\}$ --- one per inport. 
For example, for the AT model, \NAME generates the input signals for the  \texttt{Throttle} and \texttt{Brake} inports.
A signal is a function $f: \timedomain \rightarrow \real$, where $\timedomain$ is a non-singular bounded interval $[0,b]$ of \real that represents the simulation time domain of system \system.
The \simulink simulator requires engineers to specify the input of the simulation. 
\Cref{fig:InputAT} presents an example of an input for the AT model. The input contains the  two input signals for the  \texttt{Throttle} and \texttt{Brake} inports defined over the simulation time domain $[0,50]$s.

Assumption \asmpt guides the generation of new input signals. 
For each input signal $\inputsignal_i \in \inputs$, it contains a triple $\langle int_i , R_i , n_i \rangle$ made of an interpolation function ($int_i$), a value range ($R_i$) and a number of control points ($n_i$). 
\NAMESIMULINK generates each input signal $\inputsignal_i$ by selecting $n_i$ time instants, i.e., $t_1,t_2,\ldots,t_{n_i}$, within the time domain $\timedomain=[0,b]$, such that $t_1=0$, $t_{n_i}=b$ and  $t_1<t_2<\ldots<t_{n_i}$.
The values of $t_1,t_2,\ldots,t_{n_i}$ can also be chosen to ensure a fixed difference between consecutive time instants, i.e., for all $t_j,t_{j+1}$ with $j\in\{1,2,\ldots n_i-1\}$, the value of $t_{j+1}-t_{j}$ is fixed.
Then, \NAMESIMULINK selects a value $\inputsignal_i(t_j)$ from value range $R_i$ for each time instant $t_j$. 
The interpolation function (e.g., piecewise constant, linear or piecewise cubic) is then used to generate the values assumed by the input signal $\inputsignal_i$ over the rest of the time domain $\timedomain$.
For example, the input signals for the  \texttt{Throttle} and \texttt{Brake} reported in \Cref{fig:InputAT} are generated by considering the value range $[0,100]$ for the throttle percentage applied to the engine (\%), and $[0,325]$ for the pound-foot (lb-ft) torque applied by the brake. 
To generate the input signals, the Piecewise Cubic Hermite Interpolating Polynomial (\texttt{pchip}) interpolation function~\cite{PCHIP} and $7$ and $3$ control points are respectively considered for the \texttt{Throttle} and the \texttt{Brake}.
The values selected for the time instants are indicated in the \Cref{fig:InputAT} with the ``$\circ$'' symbol on the x-axis. 
The points on the input signals labeled with the ``$\ast$'' symbol indicate the values selected for the input signals at these time instants.
The input generator component assigns values to the search parameters: the values to be assigned to the control points.
 The number of search parameters is equal to the sum, across the signals, of the number of their control points.
Note that the number of possible combinations to be assigned to the parameters depends on their range and type. For example, if the values assumed by the control points are real, there are infinite possible combinations of parameter values.

\emph{System Execution} (\phase{2}).
\NAMESIMULINK uses the
\simulink simulator to execute system \system for input \inputs and to produce output \outputs, i.e., $\outputs=\system(\inputs)$. 
The output is a set  $\outputs=\{\outputsignal_1, \outputsignal_2, \ldots, \outputsignal_m\}$ of signals (a.k.a. output signals) --- one per outport. 
\Cref{fig:OutputAT} shows the output of the AT model corresponding to the input in  \Cref{fig:InputAT}.
The output is made by three output signals associated with the \texttt{Speed}, \texttt{RPM}, and \texttt{Gear} outports.

\emph{Fitness Assessment} (\phase{3}). \NAMESIMULINK computes fitness measure $\fitness(\outputs)$ associated with the output of the system execution. 
Note that since connections within the \simulink model can directly connect inports to outports, fitness measures can also use information from input signals to guide the search. \NAMESIMULINK implements the \NAME framework by enabling the computation of the fitness measure as explained in steps \phase{4} - \phase{6}.

\NAME \emph{Fitness} (\phase{4}). It is a function that combines the values computed by the manually defined and automatically generated fitness functions.
An example of a fitness function that linearly combines the values assumed by the manually defined and automatically generated fitness functions is as follows: 
\begin{align}
    \fitness=\fitness_a\cdot p+\fitness_m \cdot (1-p) 
    \label{eq:AthenaFitness}\nonumber
\end{align}
where $p$ is a parameter within the range $[0,1]$.
\NAMESIMULINK considers solely the value of the automatic fitness when $p = 1$, and solely the value of the manual fitness when $p = 0$.
The automatic fitness value is prioritized more for higher values of $p$, while the manual fitness value is prioritized more for lower values.
Engineers may set the parameter $p$ to the value $0.5$ to equally prioritize manual and automatic fitness measures when analyzing the AT model.
An analysis of how the parameter $p$ affects the performance of the framework is presented in Section \ref{sec:evaluation}.

\NAMESIMULINK can use more complex functions to combine the manual and automatic fitness measures, and it can dynamically change the fitness function during the search (e.g.,~\cite{xu2018dynamic}).
How engineers can customize the \NAME \emph{Fitness} based on their needs is discussed in the next sections.

\emph{Automatic Fitness} (\phase{5}). \NAMESIMULINK enables engineers to automatically generate fitness functions from a requirement~$\requirement$ expressed using a logic-based formalism, such as Signal Temporal Logic~(STL)~\cite{maler2004monitoring} or Restricted Signals First-Order Logic (RFOL)~\cite{menghi2019generating}. 
For example, consider requirement AT1 that specifies that the value of the \texttt{Speed} output signal shall be  lower than $120$mph for every instant within $[0,20]$ time interval.
The requirement can be expressed in STL as
\begin{center}
    $\LTLg_{[0,20]} (\texttt{Speed}<120)$,
\end{center}
where ``\texttt{Speed}<$120$'' is a predicate indicating that the ``\texttt{Speed}'' is lower than the value ``$120$'', $\LTLg$ is the ``globally'' temporal operator, and $[0,20]$ is a time interval indicating that the predicate must hold from the time instant ``$0$'' to the time instant ``$20$''.
\NAMESIMULINK  automatically translates the STL specification into a fitness function $f_a$.
The value  generated by the fitness function $f_a$ is negative if the property is violated and positive otherwise.
As before, the magnitude of the value reflects the relative distance the system is from failing or satisfying its requirement.

\emph{Manual Fitness} (\phase{6}).  \NAMESIMULINK enables engineers to define a fitness function that considers the input (\inputs) of the \simulink simulation, the model of system \system, and assumption \asmpt for the computation of fitness value $\fitness_m$. For example, given a property that requires the vehicle speed to be lower than $120$mph at all times, the function 
\begin{center}
$\fitness_m=mean(\texttt{Brake})-mean(\texttt{Throttle})$,
\end{center}
 is a possible manual fitness function for the AT model,
where $mean(\texttt{Throttle})$ and $mean(\texttt{Brake})$ are the average values assumed by the \texttt{Throttle} and \texttt{Brake} input signals over the simulation time.
The value assumed by $mean(\texttt{Brake})$
increases as the average value assumed by the input signal \texttt{Brake} increases. 
The value assumed by the term $-mean(\texttt{Throttle})$ decreases as the average value assumed by the input signal \texttt{Throttle} increases.
Since \NAME minimizes the value computed by the fitness function, the function $f_m$ guides the search toward areas of the input domain with high \texttt{Throttle} and low \texttt{Brake} values that are more likely to make the speed of the vehicle higher than $120$mph.
Specifically, the higher the value of the $mean(\texttt{Throttle})$,  the lower the value of $f_m$, and the lower the value of the $mean(\texttt{Brake})$, the lower the value of $f_m$.



\section{Implementation}
\label{sec:implementation}
We implement \NAMESIMULINK, a Matlab application publicly available online~\cite{ATheNA}.
\NAMESIMULINK is also available as a \simulink Add-On on its marketplace~\cite{ATheNAAddon}.
\NAMESIMULINK is a plugin for \staliro~\cite{S-Taliro}, an open-source SBST tool.
We selected \staliro, among other alternatives (e.g., \Breach~\cite{Breach}, 
\FalStarTOOL~\cite{ernst2019fast}, 
\FalCAuN~\cite{Falstar}, \falsify~\cite{yamagata2020falsification}, \FalStar~\cite{ernst2019fast}, \foresee~\cite{falsQBRobCAV2021}) 
due to its recent classification as ready for industrial development~\cite{kapinski2016simulation}, and its use in several industrial systems (e.g.,~\cite{tuncali2018experience}). 
In addition, this choice makes our solution applicable to other \staliro plugins, such as \Aristeo~\cite{Aristeo}.

\lstset{language=Matlab,
    breaklines=true,
    keywordstyle=\color{blue},%
    morekeywords=[2]{1}, keywordstyle=[2]{\color{black}},
    identifierstyle=\color{black},%
    stringstyle=\color{mylilas},
    commentstyle=\color{mygreen},%
    showstringspaces=false,
    numbers=left,%
    numberstyle={\tiny \color{black}},
    numbersep=9pt, 
    emph=[1]{for,end,break,Abstract,true},emphstyle=[1]\color{blue},   morekeywords={classdef,methods},
    basicstyle=\ttfamily
}

\newcommand{\indentrule}{\color{gray}\rlap{\smash{\hspace{6pt}\rule[-.35em]{1pt}{1.45em}}}}

\newcommand{\frameX}{0.20\textwidth}

\begin{figure}[t]
\begin{lstlisting}[frame=single,caption={\NAMESIMULINK  Implementation.},label={listingMatlab},escapechar=|,basicstyle=\footnotesize,numbersep=5pt,xleftmargin=\frameX,xrightmargin=\frameX]
|\label{lst:beginabs}|classdef (Abstract) F_Assessment
    methods (Abstract=true)
|\label{lst:beginabstwo}|        fa=autFitness(S,A,phi,|\inputs|,|\outputs|);
|\label{lst:beginabsone}|        fm=manFitness(S,A,phi,|\inputs|,|\outputs|);
|\label{lst:beginabsthree}|        f=athenaFitness(S,A,phi,|\inputs|,|\outputs|);
|\label{lst:stop}|        s=stopCriterion(S,A,phi,|\inputs|,|\outputs|);
    end
|\label{lst:endabs}|end

|\label{lst:beginatabs}|classdef AT1_F_Assessment < F_Assessment
    methods
|\label{lst:automatedbe}|        function fa=autFitness(S,A,phi,|\inputs|,|\outputs|)
|\label{lst:callta}|           return callTaliro(o,phi,[...]);
|\label{lst:automatedend}|        end
|\label{lst:manstart}|        function fm=manFitness(S,A,phi,|\inputs|,|\outputs|)
|\label{lst:throttlef}|           throttlef=scale(mean(|\inputs|(:,1)), A(1).R);
|\label{lst:brakef}|           brakef=scale(mean(|\inputs|(:,2)), A(2).R);
|\label{lst:ret}|           return brakef-throttlef;
|\label{lst:manend}|        end
|\label{lst:canalstart}|        function f=athenaFitness(S,A,phi,|\inputs|,|\outputs|)
           return 0.5*autFitness(S,A,phi,|\inputs|,|\outputs|)
                  +0.5*manFitness(S,A,phi,|\inputs|,|\outputs|);
|\label{lst:canalend}|        end
|\label{lst:stopbegin}|        function s=stopCriterion(S,A,phi,|\inputs|,|\outputs|)
           return autFitness(S,A,phi,|\inputs|,|\outputs|)<0;
|\label{lst:stopend}|        end
    end
|\label{lst:endatabs}|end
\end{lstlisting}
\end{figure}

\NAMESIMULINK reuses the modules provided by \staliro to implement the input generation~(\phase{1}) and system execution~(\phase{2}) steps of \NAME.
For the input generation step, \staliro provides a set of alternatives that rely on different search algorithms, such as Simulated Annealing~\cite{abbas2014robustness}, Monte Carlo~\cite{nghiem2010monte}, and gradient descent methods~\cite{abbas2014functional}.
These algorithms automatically select the values  $\inputsignal_i(t_j)$ from the value range $R_i$ for each time instant $t_j$ (see \Cref{sec:applSimulink}).
For the system execution step, \staliro relies on the \texttt{sim} command of Matlab~\cite{sim} to run the  \simulink simulator. 

\NAMESIMULINK modifies the fitness assessment step of  \staliro~(\phase{3}) as described in Section~\ref{sec:applSimulink}. 
Specifically, the source code of \staliro was modified to receive a subclass of the  \texttt{F\_Assessment} abstract class as input to compute the fitness measure. 
The abstract class \texttt{F\_Assessment} detailed in Listing~\ref{listingMatlab} (Lines~\ref{lst:beginabs}-\ref{lst:endabs}) describes the generic common functionalities of \NAMESIMULINK fitness functions.
More precisely, it specifies that each \NAMESIMULINK fitness measure has three methods: 
\texttt{autFitness} (Line~\ref{lst:beginabstwo}), that specifies how the automatic fitness measure is computed (\phase{5}),
\texttt{manFitness} (Line~\ref{lst:beginabsone}), that specifies how the manual fitness measure is computed (\phase{6}),
 and
\texttt{athenaFitness} (Line~\ref{lst:beginabsthree}), that specifies how the automatic and manual fitness values are combined (\phase{4}).
Finally, the  \texttt{stopCriterion} method defines the condition that stops the search.

\NAMESIMULINK allows for model-specific fitness functions that combine manually defined and automatically generated fitness functions. For example, the subclass \texttt{AT1\_F\_Assessment}, detailed in Listing~\ref{listingMatlab} (Lines~\ref{lst:beginatabs}-\ref{lst:endatabs}), provides the implementation for the methods of the class \texttt{F\_Assessment} for the requirement AT1 of AT.

The method \texttt{autFitness} (Lines~\ref{lst:automatedbe}-\ref{lst:automatedend}),  which computes the value of the automated fitness function, is implemented by using the method \texttt{callTaliro} provided by \staliro which supports STL specifications.
As done in the default implementation of \staliro, we provided the output signals (\outputs) generated by running the \simulink model  as inputs to the method \texttt{CallTaliro}, together with the property \texttt{phi}, and some additional configuration parameters omitted for brevity ($[...]$ in Line~\ref{lst:callta}).
A full explanation of how \staliro computes the fitness value (and the supported expressions) is outside the scope of this work and can be found in the corresponding publication~\cite{S-Taliro}.
Note that, in the literature, there exist multiple definitions of automatic fitness or robustness measures (e.g.,~\cite{fainekos2009robustness,fainekos2006robustness,Donze2010,Pant2017,menghi2019generating}) and our implementation can be extended to support these fitness functions.

Method \texttt{manFitness} (Lines~\ref{lst:manstart}-\ref{lst:manend}) implements the manual fitness function by defining variables \texttt{throttlef} (Line~\ref{lst:throttlef}) and \texttt{brakef} (Line~\ref{lst:brakef}).
The variable \texttt{throttlef} contains the average (\texttt{\textcolor{blue}{mean}(}\inputs\texttt{(:,1))})  of the values stored in the first column (\inputs\texttt{(:,1)}) of the input (\inputs), that is average of the values assumed by the \texttt{Throttle} input signal.
This value is scaled (\texttt{scale}) within $[0,1]$, by considering the value range (\texttt{A(1).R}) for the throttle.
The higher the value associated with the \texttt{Throttle}, the higher the value assumed by the variable \texttt{throttlef}.
The variable \texttt{brakef} is computed in the same way but with the second column of the input.
The manual fitness function value is the difference between the values of the variables \texttt{brakef} and \texttt{throttlef} (Line~\ref{lst:ret}), which is within the range $[-1,1]$. The value $-1$ means \texttt{throttle} is maximum and no \texttt{brake} is applied and the value $1$ indicates the opposite. 
Since the goal of the search is to minimize the fitness value, the manual fitness function ensures that input signals with high \texttt{Throttle} and low \texttt{Brake} are prioritized during the search.

The  \texttt{athenaFitness} method (Lines~\ref{lst:canalstart}-\ref{lst:canalend}), that computes the value of the \NAME fitness function, computes the mean of the values assumed by the automatic and manual fitness functions. 
Since the values of automated and manual fitness functions are within the range $[-1,1]$, this fitness function ensures that the \NAME fitness value is also within the range $[-1,1]$ and both the manual and the automated fitness functions are equally prioritized. 
\NAMESIMULINK also enables engineers to define \NAME fitness functions that go beyond the one proposed in \Cref{sec:applSimulink} by defining alternative implementations for the method \texttt{athenaFitnesss}.

The  \texttt{stopCriterion}  method (Lines~\ref{lst:stopbegin}-\ref{lst:stopend}), implementing the stopping criterion, aborts the search whenever the value computed by the automatic fitness function is lower than $0$.
This stopping criterion reflects the robustness semantics of STL~\cite{fainekos2009robustness}, i.e., a negative value indicates that the STL specification of the AT1 requirement is violated.

\begin{table*}[t]
    \centering
    \caption{Identifier (MID), description, number of blocks (\#Blocks), inports (\#Inport), outports (\#Outport),
    simulation time in seconds (Ts), requirements (\#Reqs) and average runtime$^{\ast}$ in seconds (Tr) of our benchmark models.}
    \label{tab:models}
    \footnotesize
    \begin{tabular}{l p{5.5cm} r r r r r r}
        \toprule
        \textbf{MID} &\textbf{Description} &\textbf{\#Blocks}   &\textbf{\#Inport}   &\textbf{\#Outport}
        & \textbf{Ts}   &\textbf{Tr}
        &\textbf{\#Reqs}\\
        \midrule
        AT    &A model of a car automatic transmission with gears from 1 to 4.   &69   &2   &3 & $50$ & $0.36$ & 10   \\
  
        AFC    & A  controller for the air-fuel ratio in an engine.   &302   &2   &3 & $50$ &1.14   &3      \\
    
        NN    & A Neural Network controller for a levitating magnet above an electromagnet.   &111   &1   &1 & $40$ &0.32   & 2      \\
     
        WT    &A model of a wind turbine that takes as input the wind speed.   &161   &1   &6 & $630$ &2.37 &4       \\
       
        CC    & A simulation of a system formed by five cars.  &13 &2  &5  & $100$ &1.28    &6\\
    
        F16    & Simulation of an F16 ground collision avoidance controller.   &55   &0   &4 & $15$ &2.76   &1      \\
    
        SC    & Dynamic model of steam condenser, controlled by a Recurrent Neural Network.   &172   &1   &4 & $35$ &1.55   &1     \\
        \bottomrule
    \end{tabular}
    \begin{flushleft}
        $\ast$ The average runtime is estimated using a 2021 MacBook Pro with an Apple M1 Pro chip.
    \end{flushleft}
\end{table*}

\begin{table*}[!ht]
    \caption{Identifier (RID), formal specification, and description for the requirements of the different models.}
    \label{tab:requirements}
    \footnotesize
    \begin{tabular}{p{0.5cm} p{12.7cm} }
    \toprule
       \textbf{RID}  & \textbf{STL Specification and Description} 
         \\
       \midrule
        AT1 & $\LTLg_{[0,20]} (\texttt{Speed}<120)$: The speed within the interval $[0,20]$s shall be lower than $120$mph.  \\
        AT2 & $\LTLg_{[0,10]} (\texttt{RPM}<4750)$: The motor speed within the interval $[0,10]$s shall be lower than $4750$rpm.  \\
        AT51 & $\LTLg_{[0,30]} ((\neg g1 \wedge \LTLf_{[0.001,0.1]} g1) \Rightarrow \LTLf_{[0.001,0.1} (\LTLg_{[0,2.5]} g1))$: If gear one is engaged within $[0, 30]$s, it shall remain engaged for $2.5$s.\\
        AT52 & $\LTLg_{[0,30]} ((\neg g2 \wedge \LTLf_{[0.001,0.1]} g2) \Rightarrow \LTLf_{[0.001,0.1]} (\LTLg_{[0,2.5]} g2))$: If gear two is engaged within $[0, 30]$s, it shall remain engaged for $2.5$s.\\
        AT53 & $\LTLg_{[0,30]} ((\neg g3 \wedge \LTLf_{[0.001,0.1]} g3) \Rightarrow \LTLf_{[0.001,0.1]} (\LTLg_{[0,2.5]} g3))$: If gear three is engaged within $[0, 30]$s, it shall remain engaged for $2.5$s.\\
        AT54 & $\LTLg_{[0,30]} ((\neg g4 \wedge \LTLf_{[0.001,0.1]} g4) \Rightarrow \LTLf_{[0.001,0.1]} (\LTLg_{[0,2.5]} g4))$: If gear four is engaged within $[0, 30]$s, it shall remain engaged for $2.5$s.\\
        AT6a & $(\LTLg_{[0,30]}(\texttt{RPM}<3000)) \Rightarrow (\LTLg_{[0,4]}(\texttt{Speed}<35))$:  \texttt{Speed} shall be lower than $35$ within $[0,4]$s, if \texttt{RPM} is lower than $3000$ within $[0,30]$s.\\
        AT6b & $(\LTLg_{[0,30]}(\texttt{RPM}<3000)) \Rightarrow (\LTLg_{[0,8]}(\texttt{Speed}<50))$:   \texttt{Speed} shall be lower than $50$ within $[0,8]$s, if \texttt{RPM} is lower than $3000$ within $[0,30]$s. \\
        AT6c & $(\LTLg_{[0,30]}(\texttt{RPM}<3000)) \Rightarrow (\LTLg_{[0,20]}(\texttt{Speed}<65))$:  \texttt{Speed} shall be lower than $65$ within $[0,20]$s, if \texttt{RPM} is lower than $3000$ within $[0,30]$s.\\
        AT6abc & $\text{AT6a} \wedge \text{AT6b} \wedge \text{AT6c}$: The requirements with RID \text{AT6a},  \text{AT6b}, \text{AT6c} shall be satisfied.\\
        \midrule
        AFC27 & $\LTLg_{[11,50]} ((rise \vee fall) \Rightarrow (\LTLg_{[1,5]} |\mu|< 0.008) )$: The error ($\mu$) shall be lower than $0.008$, if the throttle angle rises or falls in $[11,50]$s.$^\ast$
        \\
        AFC29 & $\LTLg_{[11,50]} (|\mu|< 0.007)$: Within $[11,50]$s,
        the error shall be lower than $0.007$.$^\dagger$ 
        \\
        AFC33 & $\LTLg_{[11,50]} (|\mu|< 0.007)$: Within $[11,50]$s, the  error shall be lower than $0.007$.$^\S$
        \\
        \midrule
        NN & 
        $\LTLg_{[1,37]}
((|Pos-Ref|>(0.005 + 0.03|Ref |)) \Rightarrow \LTLf_{[0,2]}\LTLg_{[0,1]}\neg (0.005+0.03|Ref | \leq |Pos-Ref|))$:
The discontinuities between the 
levitating magnet position ($Pos$) and the reference position ($Ref$) shall be at least 3 time units apart.
\\
        NNx &
$\LTLf_{[0,1]}(Pos > 3.2) \wedge \LTLf_{[1,1.5]}(\LTLg_{[0,0.5]}(1.75 < Pos < 2.25)) \wedge \LTLg_{[2,3]}(1.825 < Pos < 2.175)$:
The magnet position ($Pos$) shall be higher than $3.2$ within $[0,1]$s, lower than $2.175$ and higher than $1.825$ within $[2,3]$s, and higher than $1.75$ and lower than $2.25$  for $0.5$s within whithin $[2,3]$s.
        \\
        \midrule
        WT1 & 
        $\LTLg_{[30,630]} (\theta \leq 14.2)$: The pitch angle shall be smaller than $14.2$deg.
        \\
        WT2 & $\LTLg_{[30,630]}(21000 \leq M_{g,d} \leq 47500)$: The torque shall be between  $21000$ and $47500$N$\cdot$m. \\
        WT3 &  $\LTLg_{[30,630]} (\Omega \leq 14.3)$: The rotor speed shall be lower than $14.3$rpm.\\
        WT4 & $\LTLg_{[30,630]} \LTLf_{[0,5]} (|\theta-\theta_d| \leq 1.6)$: The commanded and the measured pitch angles differ for at most $1.6$deg.
        \\
        \midrule
        CC1 & $\LTLg_{[0,100]} (y_5 -y_4 \leq 40)$:  Within the interval $[0,100]s$, the difference between $y_5$ and $y_4$ shall be lower than $40$. \\
        CC2 & $\LTLg_{[0,70]} \LTLf_{[0,30]} (y_5-y_4 \geq 15)$: For every instant in $[0,70]s$, 
        the value of $y_5-y_4$ shall exceed $15$ 
        in one instant within the next $30$s.
        \\
        CC3 & 
     $\LTLg_{[0,80]}((\LTLg_{[0,20]} (y_2-y_1\leq20))\vee (\LTLf_{[0,20]} (y_5-y_4 \geq 40)))$: For every instant in $[0,70]s$, 
      either $y_2-y_1$ shall be lower than $20$ for the next $20$s or the value of $y_5-y_4$ shall be higher than $40$ for the next $20$s.
        \\
        CC4 & $\LTLg_{[0,65]} \LTLf_{[0,30]} 
        \LTLg_{[0,20]}(y_5-y_4 \geq 8)$:
        For every instant in $[0,65]$s,
        within the next $30$s, 
     $y_5-y_4$ shall be higher than $8$ for at least $20$s.
        \\
        CC5 & $\LTLg_{[0,72]} \LTLf_{[0,8]} (
        (\LTLg_{[0,5]} (y_2-y_1 \geq 9))
        \Rightarrow
        (\LTLg_{[5,20]}( y_5-y_4 \geq 9))
        )$:
        For every time instant in $[0,72]$s,
         if within the next $8$s the value of $y_2-y_1$ is higher than $9$ for at least $5$s,
        after $5$s the value of $y_5-y_4$ shall be higher than $9$ and remain higher than $9$ for the following $15$s.
        \\
        CCx & $\bigwedge_{i=1..4} \LTLg_{[0,50]} (y_{i+1}-y_{i}>7.5)$: 
        The difference between $y_{i+1}$ and $y_{i}$ shall be higher than $7.5$ within $[0,50]s$.  \\
        \midrule
        F16 & $\LTLg_{[0,15]} (altitude>0)$:  Within the interval $[0,15]s$, the altitude shall be higher than $0$. \\
    \midrule
        SC & $\LTLg_{[30,35]} (87 \leq pressure \leq 87.5)$: The pressure shall remain between $87.5$ and $87$ within  $[30,35]s$. \\
    \bottomrule
    \end{tabular}
    \begin{flushleft}
    $\ast$ $rise=(\theta<8.8) \wedge (\LTLf_{[0,0.005]} (\theta>40.0))$, \hspace{0.3cm}
    $fall=(\theta>40.0) \wedge (\LTLf_{[0,0.005]} (\theta<8.8))$, \hspace{0.3cm}
    $ 0 \leq \theta<61.2$\\
    $\dagger$ The range of the throttle angle is $0 \leq \theta<61.2$.\\
    $\S$ The range of the throttle angle is $61.2 \leq \theta \leq 81.2$.
    \end{flushleft}
\end{table*}


\section{Evaluation}
\label{sec:evaluation}
In this section, we empirically evaluate \NAME by (a)~comparing the effectiveness and efficiency of \NAMESIMULINK with that of an existing SBST framework for \simulink models (\textbf{RQ3} and \textbf{RQ4}) and (b)~assessing its usefulness on two large and representative case studies from the literature (\textbf{RQ5}).
However, to answer these questions, we first must select the manual fitness functions and the \NAME fitness functions to be considered in our experiments.
Therefore, we first answer two research questions that identify the manual fitness function (\textbf{RQ1}) and the \NAME fitness function (\textbf{RQ2}) to be considered in our experiments.

\vspace{0.3cm}
\emph{Configuration of \NAME.} To select the configuration of \NAME to be used in our experiments, we consider the following research questions:\\
$\bullet$ \textbf{Manual Fitness Function Design - RQ1.} \emph{How complex is it to write the manual fitness functions?} (Section~\ref{sec:manualfitnessinfluence})\\
We reverse engineer a set of benchmark \simulink models, propose a set of manual fitness functions, and check their effectiveness in generating failure-revealing test cases. We also assess how complex it is for engineers to write effective manual fitness functions.\\ 
$\bullet$ \textbf{\NAME Fitness Selection - RQ2.} \emph{How does the selection of the \NAME fitness influence the generation of failure-revealing test cases? 
How do we determine what are the optimal values for the parameter $p$ for the \NAME fitness function from \Cref{sec:applSimulink}?} (Section~\ref{sec:athenafitnessinfluence})\\
We compare the effectiveness and efficiency of \NAMESIMULINK for different fitness functions. 
We consider the manual fitness functions designed for RQ1 and the \NAME fitness function from \Cref{sec:applSimulink},
and analyze how different values for the parameter~$p$ influence the generation of failure-revealing test cases. 
We identify the optimal values for the parameter $p$ for our benchmark models.

\vspace{0.3cm}
\emph{Comparison with existing SBST frameworks:} To compare the effectiveness and efficiency of \NAMESIMULINK with the one of an existing SBST framework for \simulink models, we consider the following research questions:\\
\noindent $\bullet$ \textbf{Effectiveness - RQ3.} \emph{How effective is \NAMESIMULINK in generating failure-revealing test cases?} (Section~\ref{sec:rq1})\\
We use \NAMESIMULINK  with the optimal configuration identified in RQ2 and evaluate its effectiveness by comparing it with existing SBST frameworks that only support automatic or manual fitness functions.
As a baseline framework supporting automatic fitness functions, we consider \NAMEAUTOMATIC, a version of \NAME obtained by considering the function $f$ presented in Section~\ref{sec:applSimulink} and by setting $0$ as value for the parameter~$p$; that is, it only uses the value computed by the automatic fitness function. 
This instance corresponds to \staliro.
We could not identify a baseline SBST framework for manual fitness functions since (a)~SBST frameworks that rely on manual fitness functions are generally problem-specific, (b)~we are not aware of a generic SBST framework based on manual fitness functions for \simulink models. 
Therefore, as a baseline framework supporting manual fitness functions, we consider \NAMEMANUAL,  a version of \NAME obtained by setting $1$ as the value for the parameter~$p$; that is, it only uses the value computed by the manual fitness function. 
We compared the capability of each tool in generating failure-revealing test cases.

\noindent $\bullet$ \textbf{Efficiency - RQ4.} \emph{How efficient is \NAMESIMULINK in generating failure-revealing test cases?}\\
We use \NAMESIMULINK  with the optimal configuration identified in RQ2 and evaluate its efficiency by comparing it with the SBST frameworks that only support automatic or manual fitness functions.
To assess the efficiency of the tools, we compare the number of search iterations required by each tool to generate failure-revealing test cases.

\vspace{0.3cm}
\emph{Assessment of usefulness:} To assess the usefulness of \NAMESIMULINK on large and representative case studies from the literature, we consider the following research question:\\
\noindent $\bullet$ \textbf{Usefulness - RQ5.}
\emph{How applicable and useful is \NAMESIMULINK in generating failure-revealing test cases for two large and representative \simulink models?} (Section~\ref{sec:rq3})\\
To assess the applicability of \NAMESIMULINK, we evaluate its effectiveness and efficiency on two large and representative case studies from the automotive and medical domains.
For the automotive case study, we inject a fault in the model.
We evaluate if \NAMESIMULINK can generate a failure-revealing test case for our two case studies.
The goal of this question is not to compare \NAMESIMULINK  with other tools,
so we did not compare \NAMESIMULINK  with \NAMEAUTOMATIC and \NAMEMANUAL. 

We make our (sanitized) models, data, and tool available online~\cite{ATheNA}.
We run our experiments on a large computing platform.\footnote{1109 nodes, 64 cores, memory 249G or 2057500M, CPU 2 x AMD Rome 7532 \@ 2.40 GHz 256M cache L3}
In the following section, we first present the benchmark models for comparing \NAMESIMULINK with existing SBST frameworks and then discuss each research question.

\subsection{Benchmark}  
Our benchmark consists of the models of the ARCH competition~\cite{DBLP:conf/arch/ErnstABCDFFG0KM21} – an international competition among testing tools for continuous and hybrid systems~\cite{ARCHWEBSITE}.
This benchmark consists of seven models: Automatic Transmission (AT), Fuel Control on Automotive Powertrain (AFC), Neural Network Controller (NN), Wind Turbine (WT), Chasing Cars (CC), Aircraft Ground Collision Avoidance System (F16), and Steam Condenser (SC).
For each model, Table~\ref{tab:models} contains a model identifier (MID), a short description of the model, the number of \simulink blocks (\#Blocks), inports (\#Inport), outports (\#Outport), simulation time (Ts), average runtime per iteration (Tr), and the number of requirements (\#Reqs).
The number of blocks, inports, and outputs varies across the models.
The models are representative: they come from different domains, including the automotive  (AT, AFC), neural networks (NN), and energy (SC) domains. They also include both discrete (e.g., logic decisions and state machines) and continuous (e.g., dynamical systems) behaviors.
The benchmark models also include AFC~\cite{jin2014powertrain}, a model developed by Toyota.

The models of the ARCH competition are associated with the $27$ requirements presented in Table~\ref{tab:requirements}.
Table~\ref{tab:requirements} presents the STL specification and a short description for each requirement.
Requirements are associated with a requirement identifier (RID) that starts with the identifier of the model. 
For example, the requirement with the identifier AT54 refers to the AT model.
The symbols ``$\LTLg$'' and ``$\LTLf$'' used in the STL specification represent the ``globally'' and ``eventually'' temporal operators.
The ``globally'' temporal operator is discussed in Section~\ref{sec:applSimulink}.
The ``eventually'' temporal operator is labeled with a subscript containing the time interval to consider when assessing the operator. 
The operator scopes a condition that shall eventually hold within the time interval. 
For example, ``$\LTLf_{[0,10]} x<5$'' indicates that the value of $x$ shall be lower than $5$ at some point within the interval $[0,10]$s.
The requirements of our benchmark have a different structure and use various temporal operators.
Out of \numrequirements requirements, only two requirements (CC1, F16) are invariants: assertions that must hold during the entire simulation of the model. 
Other requirements scoped with the $\LTLg$ operator (e.g., AT1) are not invariants since the time bound of the operator (e.g., $[0,20]$s) does not force the requirement to hold at every time instant during the simulation (e.g., $50$s).

\begin{table*}[]
    \caption{Interpolation function ($int$), value range ($R$ and $R^\prime$), and number of control points ($n$) for the input signals.}
    \label{tab:assumptions}
    \centering
    \footnotesize
    \begin{tabular}{l l p{4cm} p{4cm} l}
    \toprule
        \textbf{RID} & $\mathbf{int}$ & $\mathbf{R}$ &  $\mathbf{R}^\prime$ & $\mathbf{n}$ \\
        \midrule
        AT1  &\texttt{pchip},\texttt{pchip} & $[0,100],[0,325]$ & $[0,110],[0,100]$ & $7, 3$ \\
        AT2 &\texttt{pchip},\texttt{pchip} &$ [0,100],[0,325]$ & $[0,90],[32,325]$ & $7, 3$\\ 
        AT51 &\texttt{pchip},\texttt{pchip} & $[0,100],[0,325]$ & & $7,3$\\
        AT52 &\texttt{pchip},\texttt{pchip} &$ [0,100],[0,325]$ & & $7,3$\\
        AT53 &\texttt{pchip},\texttt{pchip} &$[0,100],[0,325]$ & & $7,3$\\
        AT54 &\texttt{pchip},\texttt{pchip} & $[0,100],[0,325]$ & & $7,3$\\
        AT6a &\texttt{pchip},\texttt{pchip} & $[0,100],[0,325]$ & $[0,47],[172,325]$ & $7,3$\\
        AT6b &\texttt{pchip},\texttt{pchip} & $[0,100],[0,325]$ & $[0,48],[169,325]$ & $7,3$\\ 
        AT6c &\texttt{pchip},\texttt{pchip} & $[0,100],[0,325]$ & $[0,44],[182,325]$ & $7,3$\\
        AT6abc &\texttt{pchip},\texttt{pchip} & $[0,100],[0,325]$ & $[0,44],[182,325]$ & $7,3$\\
        \midrule
        AFC27 & \texttt{const},\texttt{pconst} & $[900,1100],[0,61.2]$ & & $1,10$\\
        AFC29 & \texttt{const},\texttt{pconst} & $[900,1100],[0,61.2]$ & &  $1,10$ \\
        AFC33 & \texttt{const},\texttt{pconst} & $ [900,1100],[61.2,81.2]$ & & $1,10$ \\
        \midrule
        NN &\texttt{pchip} & $[1,3]$ & $[1,2]$ & $3$\\
        NNx &\texttt{pchip} & $[1.95,2.05]$ & $[1.95,2.14]$ & $3$\\
        \midrule
        WT1 &\texttt{pchip} & $[8,16]$  & $[7.5,16.5]$ & $126$\\
        WT2 &\texttt{pchip} & $[8,16] $ & & $126$\\
        WT3 &\texttt{pchip} & $[8,16] $ & & $126$\\
        WT4 &\texttt{pchip} & $ [8,16]$ & & $126$\\
        \midrule
        CC1 &\texttt{pchip},\texttt{pchip} & $ [0,1],[0,1] $ &$[0,0.82],[0.18,1]$ & $7,3$ \\
        CC2 &\texttt{pchip},\texttt{pchip} & $ [0,1],[0,1]$  & & $7,3$ \\
        CC3 &\texttt{pchip},\texttt{pchip} & $ [0,1],[0,1]$ & & $7,3$ \\
        CC4 &\texttt{pchip},\texttt{pchip} & $ [0,1],[0,1]$ & & $7,3$ \\
        CC5 &\texttt{pchip},\texttt{pchip} & $ [0,1],[0,1]$ & & $7,3$ \\
        CCx &\texttt{pchip},\texttt{pchip} & $ [0,1],[0,1]$ & & $7,3$ \\
        \midrule
        F16 
        & \texttt{const},\texttt{const},\texttt{const} &  $[0.63,0.89],[-1.26,-1.10],\newline [-1.18,-0.39]$ & $[0.16,0.89],[-1.26,-0.63],\newline [-1.83,-0.39]$ & $1,1,1$ \\
        \midrule
        SC &\texttt{pchip} & $[3.99,4.01]$ & $[3.984,4.016]$ & $20$ \\
     \bottomrule
    \end{tabular}
        \begin{flushleft}
\hspace{0.4cm}    \texttt{pchip}: piecewise cubic,\hspace{0.5cm} \texttt{const}: constant signal,\hspace{0.5cm} \texttt{pconst}: piecewise constant signal.
    \end{flushleft}
\end{table*}

For each requirement, Table~\ref{tab:assumptions} presents the assumptions considered in the ARCH competition that we use for generating the test cases.
Specifically, for each requirement of our benchmark, the table reports the interpolation function ($int$), value range ($R$ and $R^\prime$), and the number of control points ($n$) considered for generating the test input signals.
Comma-separated values are related to different input signals.
For example, for AT1, $[0,100]$ and $[0,325]$ are the value ranges considered for generating the \texttt{Throttle} and the \texttt{Brake} input signals, respectively.
Note that while the configuration of the ARCH competition includes the input ranges, it does not include the number of control points and the interpolation functions, as different SBST frameworks can use different strategies to generate the input signals.
Thus, we select the setting used by \Aristeo in the last competition since the complete replication package is publicly available~\cite{ARIsTEOWeb}. 
For some of the requirements for which the tools were showing a similar behavior considering the value range $R$ of the ARCH~2021 competition (see the results of \emph{RQ3} and \emph{RQ4}), we consider an additional value range $R^\prime$ that enables a broader comparison between the tools.
We remark that, since control points can assume real values, for our experiments, there are infinite possible combinations of parameter values for the search algorithm to be assessed.  

Table~\ref{tab:assumptions} leads to \numrequirementassumption assumption-requirement combinations.
Each combination consists of a requirement and an assumption generated by considering the interpolation function, the number of control points, and one of the value ranges specified for that requirement. 
For example, two combinations are present for requirement AT1, generated by considering value ranges ~$R$ and~$R^\prime$.
For each assumption-requirement combination of our benchmark, the requirement can be violated.
In our evaluation, we compare the behavior of the different tools by considering each assumption-requirement combination. 
We considered the number of iterations as metric for our comparison since the time-per-iteration of the tools we considered in our experiments reduces to the time required to simulate the model. The computation time required to generate the tests and compute value of the manual fitness functions is negligible compared to the simulation time.

\begin{table*}[]
    \caption{Manual fitness functions description for our benchmark requirements.}
    \label{tab:inputs}
    \centering
    \footnotesize
    \begin{tabular}{l p{12cm} }
        \toprule
        \textbf{RID} &  \textbf{Manual Fitness Description} \\
         \midrule
        AT1  & Maximizes the lowest throttle value within $[0,17]$s and minimizes the highest brake value within $[0,25]$s.\\
        AT2   & Maximizes the average throttle value within $[0,8]$s, then minimizes the average brake value within $[0,25]$s. \\
        AT51    & Makes the first three throttle control points get as close as possible to $\{35 \%, 0 \%, 50 \%\}$ respectively and maximize brake within $[0,25]$s. \\
        AT52    & Maximizes the minimum throttle value between $[0,8]$s. \\
        AT53    & Makes the first three throttle control points get as close as possible to $\{100 \%, 20 \%, 0 \%\}$ respectively and minimize brake within $[0,25]$s. \\
        AT54    &  Makes the first three throttle control points form an upward arc and the brake ones a downward arc.\\
        AT6a &  Makes the average throttle value within $[0,33]$s as close as possible to $45 \%$ and minimizes the average brake value within $[0,25]$s. \\
        AT6b  & Makes the average throttle value within $[0,33]$s as close as possible to $45 \%$ and minimizes the average brake value within $[0,25]$s. \\
        AT6c & Makes the average throttle value within $[0,33]$s as close as possible to $45 \%$ and minimizes the average brake value within $[0,25]$s. \\
        AT6abc & Makes the average throttle value within $[0,33]$s as close as possible to $45 \%$ and minimizes the average brake value within $[0,25]$s. \\
        \midrule
        AFC27 & Increases the two control points adjacent to the lowest one  above $40 \deg$, then minimizes the lowest value within $[10,50]$s.\\
        AFC29 & Minimizes the lowest throttle value within $[10,50]$s.\\
        AFC33 & Minimizes the engine speed value.\\
        \midrule
        NN &  Minimizes the reference position control point at $20$s.\\
        NNx &  Maximizes the lowest reference position within $[0,20]$s. \\
        \midrule
        WT1 & Maximizes the steepest positive slope between two consecutive control points within $[30,630]$s.\\
        WT2 & Maximizes the steepest negative slope between two consecutive control points within $[30,630]$s.\\
        WT3 & Maximizes the steepest positive slope between two consecutive control points within $[30,630]$s.\\
        WT4 & Maximizes the average distance between consecutive control points within $[30,630]$s.\\
        \midrule
        CC1 & Maximizes the lowest throttle value within $[0,100]$s and minimizes the highest brake value within $[0,100]$s. \\
        CC2 &  Minimizes the highest throttle value within $[0,100]$s and maximizes the lowest brake value within $[0,100]$s. \\
        CC3 & Maximizes the lowest throttle value within $[0,100]$s and minimizes the highest brake value within $[0,100]$s. \\
        CC4 & Minimizes the minimum distance between cars 4 and 5 within $[0,100]$s.\\
        CC5 & Makes the average throttle value within $[0,33]$s as close as possible to 0.3 and maximizes the average brake value within $[0,50]$s. \\
        CCx &  Maximizes the throttle control point at $0s$ and minimizes the throttle control point at $17s$.\\
        \midrule
        F16 &  Maximizes the initial roll angle and minimizes the initial pitch angle. \\
        \midrule
        SC &  Maximizes the peak-to-peak distance of the steam flow rate within $[29.5,35]$s. \\
        \bottomrule
    \end{tabular}
\end{table*}

\subsection{Manual Fitness Function Design  --- RQ1}
\label{sec:manualfitnessinfluence}
To assess how complex it is to write the manual fitness functions, we reverse engineer our benchmark \simulink models, propose a set of manual fitness functions, and check their effectiveness in generating failure-revealing test cases.
Then, we assess how hard it is for engineers to write these manual fitness functions.

\textbf{Design of the manual fitness function}: 
The first author of this paper conducted the reverse engineering activity by reading the publication related to each model of our benchmark (if available online), opening the model, and looking at the structure (its \simulink blocks and how they are connected), and running the model by considering a set of inputs (approximately ten different inputs for each model).
The manual inspection required approximately five hours for each model. 
Therefore, although we have a general understanding of the model, our knowledge of the model is limited and lower than the one engineers will have in practical applications when using \NAMESIMULINK.
Then, for each requirement, we designed a manual fitness function.
We organized a set of meetings in which the first author presented the model functionalities to the other authors, and we then designed a manual fitness function that tries to guide the search toward inputs that we believed were more critical. 
The explanation of the model functionalities took approximately $10$ minutes and the formulation of the manual fitness function took around $5$ minutes for each requirement.
When formulating the manual fitness functions, we did not know the reason for the failure — we only considered the requirement specification and some general understanding of the model behavior that came from the reverse engineering activity. 
Our manual fitness functions are in \Cref{tab:inputs}.
Our manual fitness functions (\Cref{tab:inputs}) are not subsumed and are significantly different from the automatically generated ones. 
For example,  for the AT1 requirement, we design a fitness function that maximizes the value of \texttt{Throttle} and minimizes the value of \texttt{Brake} (\Cref{tab:inputs}), while the requirement requires the \texttt{Speed} to be lower than $120$ (see \Cref{tab:requirements}),

We then checked if our manual fitness functions were effective in generating failure-revealing test cases.
For each assumption-requirement combination, we run \NAMEMANUAL, a version of \NAMESIMULINK that only uses the manual fitness function (i.e., a version of \NAME obtained by considering the function $f$ from Section~\ref{sec:applSimulink} and by assigning the value $0$ to the parameter~$p$).
We run each experiment \nruns times to consider the stochastic nature of the algorithm and set the maximum number of iterations to 300, as done in similar works (e.g.,~\cite{Aristeo}) and mandated by the ARCH competition~\cite{DBLP:conf/arch/ErnstABCDFFG0KM21}.

\Cref{tab:resultsManualDesign} presents the percentage of failure-revealing runs, defined as the percentage of runs that return a failure-revealing test case (over the total number of runs), for each assumption-requirement combination from \Cref{tab:assumptions}.
The results show that for $90\%$ of the combinations ($35$ out of $39$), \NAMESIMULINK returned at least one failure-revealing run for our manual fitness functions.
Therefore, we could design an effective manual fitness function with limited knowledge about the models for most combinations.
Note that the goal of this work is not to support engineers in writing effective manual fitness functions, but to propose a framework that combines fitness functions automatically generated from requirements specifications and manually defined by engineers. 
We assess this capability in RQ3, RQ4, and RQ5.

While it is possible to address a search problem by manually defining ad-hoc fitness functions in theory, our results show that this is also possible in practice for most of the assumption-requirement combinations we considered. 
We do not know if the manual fitness functions we designed are the most effective, or if more effective manual fitness functions exist: We are not the developers of the models, and our knowledge of the model is limited and lower than the one engineers will have for industrial applications. 
Therefore, in practice, engineers will likely be more knowledgeable about their models and able to design more effective manual fitness functions.
To assess to what extent it is effective for engineers to embrace this approach, that is, how hard it is to write manual fitness functions, we proceeded as follows. 

\begin{table*}
    \centering
    \footnotesize
    \caption{Percentage of failure-revealing runs detected by using our manual fitness functions for each assumption-requirement combination of Table~\ref{tab:assumptions}. In bold, all the instances where the manual fitness function was not able to falsify the benchmark in any run.}
    \label{tab:resultsManualDesign}
    \begin{tabular}{l  r r  }
    \toprule
        \textbf{RID} & $\mathbf{R}$  &  $\mathbf{R}^{\prime}$  \\
    \cmidrule(r){1-1} \cmidrule(r){2-2} \cmidrule(r){3-3} 
        AT1     & 2 \%      & 78 \%  \\
        AT2     & 100 \%    & 88 \% \\
        AT51    & 36 \%     & \\
        AT52    & 100 \%    & \\
        AT53    & 94 \%     & \\
        AT54    & 62 \%     & \\
        AT6a    & 40 \%     & 36 \% \\
        AT6b    & 34 \%     & 32 \% \\
        AT6c    & 54 \%     & 32 \% \\
        AT6abc  & 46 \%     & 32 \% \\
    \bottomrule
    \end{tabular}
    \hspace{0.5cm}
    \begin{tabular}{l  r r  }
    \toprule
        \textbf{RID} & $\mathbf{R}$  &  $\mathbf{R}^{\prime}$ \\
    \cmidrule(r){1-1} \cmidrule(r){2-2} \cmidrule(r){3-3} 
        AFC27   & 52 \%     & \\
        AFC29   & 100 \%    & \\
        AFC33   & \textbf{0 \%}      & \\
    \cmidrule(r){1-1} \cmidrule(r){2-2} \cmidrule(r){3-3} 
        NN      & 68 \%     & 86 \% \\
        NNx     & \textbf{0 \%}      & 100 \%  \\
    \cmidrule(r){1-1} \cmidrule(r){2-2} \cmidrule(r){3-3} 
        WT1     & 2 \%      & 94 \% \\
        WT2     & 92 \%     & \\
        WT3     & 82 \%     & \\
        WT4     & 42 \%     & \\
    \bottomrule
    \end{tabular}
    \hspace{0.5cm}
    \begin{tabular}{l r r }
    \toprule
        \textbf{RID} & $\mathbf{R}$  &  $\mathbf{R}^{\prime}$  \\
    \cmidrule(r){1-1} \cmidrule(r){2-2} \cmidrule(r){3-3} 
        CC1     & 98 \%     & 6 \% \\
        CC2     & 92 \%     & \\
        CC3     & 88 \%     & \\
        CC4     & \textbf{0 \%}      & \\
        CC5     & 98 \%     & \\
        CCX     & 42 \%     & \\
      \cmidrule(r){1-1} \cmidrule(r){2-2} \cmidrule(r){3-3}
        F16     & 100 \%    & 98 \% \\
       \cmidrule(r){1-1} \cmidrule(r){2-2} \cmidrule(r){3-3}
        SC      & \textbf{0 \%}      & 30 \% \\
    \bottomrule
    \\
    \end{tabular}
\end{table*}

\textbf{Assessment}: We assess our approach with two subjects. 
Our study subjects are two bachelor students majoring in Mechatronics (Subject~1 and Subject~2) of McMaster University.
The two students have academic knowledge of Matlab/\simulink  and SBST techniques.
To assess how hard it is to write manual fitness functions, we conducted the following experiment:\\
First, we had a briefing session of $20$ minutes where we explained to them the goal of the experiments, and provided a general overview on \NAME.
Then, we selected one assumption-requirement combination for each model.
The rows of \Cref{sec:empiricalresult} report the assumption-requirement combinations we considered in these experiments.
We provided the students with a short textual description of the functionality of each model and the assumption and requirement we selected.
We asked them to propose a manual fitness function.
The students had 15 minutes to write the pseudocode of their manual fitness function.
We collected the pseudocode of the manual fitness functions designed by the students and translated them into an \NAMESIMULINK manual fitness function.
Then, we considered each manual fitness function and executed one run of \NAMEMANUAL to verify if the manual fitness functions the students proposed returned any failure-revealing run.
In this case, we set the maximum number of iterations to 1500, since we executed a single run.
We make our textual description and the fitness functions proposed by the students publicly available~\cite{ATheNA} for experiment replication.

\Cref{sec:empiricalresult} reports our results. 
Our results show that, on average, the students took approximately $6.1$min to design their manual fitness functions.
For $83\%$ ($5$ out of $6$) of our assumption-requirement combinations, at least one of the students proposed a manual fitness function that could return a failure-revealing run for one run of \NAMESIMULINK.
We conclude that the students could design an effective manual fitness function for most of the assumption-requirement combinations we considered ($83\%$). 
Note that 
(a)~the students have limited prior experience with Matlab/\simulink and with \NAME,
(b)~the students did not have any prior knowledge about the model, requirements, and assumptions other than the short textual description we provided them, and 
(c)~our textual description only covers the functionality of the models and does not provide any detail or hint about the manual fitness function.
Nevertheless, our study subjects were able to design effective manual fitness functions within $15$ minutes.
This experiment suggests that it is effective for engineers to embrace our approach - professional engineers have far more experience than our study subjects and, in practice, have more knowledge about the model, assumptions, and requirements than what we provided to our study subjects.
Our results also show that our approach was cost-effective for our assumption-requirement combinations: the students required on average $6.1$min to propose their manual fitness function.
In practice, developing industrial \simulink models requires months or even years~\cite{boll2021characteristics}.
We believe that more extensive studies, involving a greater number of participants will confirm our findings.
\begin{table}[t]
    \centering
    \footnotesize
    \caption{Result of experiment on Manual fitness functions defined by test subjects 1 
    and 2.
    }
    \label{sec:empiricalresult}
    \begin{tabular}{l l r c r  r c r}
        \toprule
            &   &\multicolumn{3}{c}{\textbf{Subject 1}}    &\multicolumn{3}{c}{\textbf{Subject 2}}\\
        \cmidrule(r){1-2}  \cmidrule(r){3-5} \cmidrule(r){6-8}
        \textbf{RID}   & \textbf{Range}    & \textbf{Time [min]}  & \textbf{Failure [Y/N]}    & \textbf{Iterations}    & \textbf{Time [min]}  & \textbf{Failure [Y/N]}   & \textbf{Iterations}\\
        \cmidrule(r){1-2}  \cmidrule(r){3-5} \cmidrule(r){6-8}
        AT1     &R$^{\prime}$   &2.0    &Y  &278    &4.0    &N  &-\\
        AFC29   &R              &6.5    &Y  &19     &6.0    &Y  &61\\
         NN      &R              &7.0    &N  &-      &5.5    &Y  &20\\
          WT3     &R              &5.5    &Y  &175    &5.5    &Y  &112\\
        CCX     &R              &13.0   &N  &-      &6.0    &N  &-\\
        SC      &R$^{\prime}$   &6.0    &Y  &1058   &6.0    &Y  &111\\
        \bottomrule
    \end{tabular}
    \label{tab:manual_fit}
\end{table}

\begin{Answer}[RQ1 - Manual Fitness Function Design]
For $90\%$ of our assumption-requirement combinations, we could design an effective manual fitness function that returned a failure-revealing test case in at least one of the $50$ runs we executed. 
Furthermore, two subjects with no prior experience with \NAME could write an effective manual fitness function for $5$ out of $6$ ($83\%$) assumption-requirement combinations they analyzed, provided with only a short description of the model functionality and its assumption and requirement.
\end{Answer}

\subsection{Influence of \NAME Fitness --- RQ2}
\label{sec:athenafitnessinfluence}
To assess how the selection of the \NAME fitness function influences the generation of failure-revealing test cases and identify the optimal values for the parameter $p$ of the \NAME fitness function for our benchmark models, we proceed as follows.

\textbf{Software Configuration:} We configure \NAMESIMULINK to use Simulated Annealing since it is the default search algorithm of \staliro.
We set the value \niterations for the maximum number of iterations since this is the value considered by the ARCH competition. 
We consider the manual fitness functions identified in RQ1.

\textbf{Methodology}: We execute \NAMESIMULINK for  each of the \numrequirementassumption assumption-requirement combinations  of \Cref{tab:assumptions}. 
To assess how the selection of the \NAME fitness function influences the generation of failure-revealing test cases, we consider different values assigned to the parameter $p$ of the \NAME fitness function (see \Cref{sec:applSimulink}).
Specifically, we run our experiments for $p$ having values of $0$, $0.2$, $0.4$, $0.5$, $0.6$, $0.8$, and $1$. 
For each combination, we run each experiment \nruns times to consider the stochastic nature of the algorithm.
For each tool, 
we record which of the \nruns runs is a failure-revealing run, i.e., it returns a failure-revealing test case.

\begin{table*}[t]
    \centering
    \footnotesize
    \caption{Percentage of failure-revealing runs for each value of $p$ and assumption-requirement combination of Table~\ref{tab:assumptions}. 
    The value of $p$ for which the assumption-requirement has the highest failure-revealing rate is highlighted in bold.}
    \label{tab:selectionOfThePValue}
    \scalebox{0.9}{
    \begin{tabular}{l r r r r r r r  r r r r r r r}
        \toprule
        &\multicolumn{7}{c}{$\mathbf{R}$}    &\multicolumn{7}{c}{$\mathbf{R}^{\prime}$}\\
        \cmidrule(r){1-1}  \cmidrule(r){2-8} \cmidrule(r){9-15}
        $\boldsymbol{p}$   &\textbf{0}  &\textbf{0.2}    &\textbf{0.4}    &\textbf{0.5}    &\textbf{0.6}    &\textbf{0.8}    &\textbf{1}  &\textbf{0}  &\textbf{0.2}    &\textbf{0.4}    &\textbf{0.5}    &\textbf{0.6}    &\textbf{0.8}    &\textbf{1}\\
        \cmidrule(r){1-1}  \cmidrule(r){2-8} \cmidrule(r){9-15}
        AT1 &\textbf{2 \%}   &\textbf{2 \%}   &\textbf{2 \%}    &\textbf{2 \%}   &0 \%   &0 \%   &0 \%
        &\textbf{78 \%}  &74 \%  &74 \%  &\textbf{78 \%}  &76 \%  &70 \%  &18 \%\\
        AT2 &\textbf{100 \%}   &98 \%   &\textbf{100 \%}   &\textbf{100 \%}   &\textbf{100 \%}   &\textbf{100 \%}   &\textbf{100 \%}
        &88 \%  &\textbf{96 \%}  &\textbf{96 \%}  &92 \%  &\textbf{96 \%}  &94 \%  &92 \%\\
        AT51 &\textbf{36 \%}   &16 \%   &20 \%   &12 \%   &14 \%   &8 \%   &4 \%\\
        AT52 &\textbf{100 \%}   &\textbf{100 \%}   &\textbf{100 \%}   &\textbf{100 \%}   &\textbf{100 \%}   &\textbf{100 \%}   &\textbf{100 \%}\\
        AT53 &94 \%   &96 \%   &\textbf{98 \%}   &94 \%   &86 \%   &88 \%   &84 \%\\
        AT54 &\textbf{62 \%}   &54 \%   &58 \%   &50 \%   &50 \%   &22 \%   &4 \%\\
        AT6a &40 \%   &82 \%   &98 \%   &96 \%   &\textbf{100 \%}   &98 \%   &\textbf{100 \%}
        &36 \%  &54 \%  &74 \%  &\textbf{86 \%}  &80 \%  &64 \%  &58 \%\\
        AT6b &34 \%   &56 \%   &\textbf{96 \%}   &92 \%   &88 \%   &\textbf{96 \%}   &88 \%
        &32 \%  &42 \%  &46 \%  &\textbf{56 \%}  &54 \%  &44 \%  &38 \%\\
        AT6c &54 \%   &84 \%   &88 \%   &90 \%   &90 \%   &\textbf{100 \%}   &90 \%
        &\textbf{32 \%}  &24 \%  &18 \%  &18 \%  &20 \%  &18 \%  &6 \%\\
        AT6abc &46 \%   &82 \%   &80 \%   &\textbf{98 \%}   &90 \%   &96 \%   &96 \%
        &\textbf{32 \%}  &24 \%  &18 \%  &18 \%  &20 \%  &18 \%  &6 \%\\
        \cmidrule(r){1-1}  \cmidrule(r){2-8} \cmidrule(r){9-15}
        AFC27 &\textbf{52 \%}   &\textbf{52 \%}   &\textbf{52 \%}   &48 \%   &50 \%   &44 \%   &38 \%\\
        AFC29 &\textbf{100 \%}   &\textbf{100 \%}   &\textbf{100 \%}   &\textbf{100 \%}   &\textbf{100 \%}   &\textbf{100 \%}   &\textbf{100 \%}\\
        AFC33 &\textbf{0 \%}   &\textbf{0 \%}   &\textbf{0 \%}   &\textbf{0 \%}   &\textbf{0 \%}   &\textbf{0 \%}   &\textbf{0 \%}\\
        \cmidrule(r){1-1}  \cmidrule(r){2-8} \cmidrule(r){9-15}
        NN &68 \%   &\textbf{74 \%}   &72 \%   &68 \%   &66 \%   &72 \%   &\textbf{74 \%}
        &86 \%  &82 \%  &86 \%  &82 \%  &\textbf{90 \%}  &80 \%  &78 \%\\
        NNx &\textbf{0 \%}   &\textbf{0 \%}   &\textbf{0 \%}   &\textbf{0 \%}   &\textbf{0 \%}   &\textbf{0 \%}   &\textbf{0 \%}
        &\textbf{100 \%}   &\textbf{100 \%}   &\textbf{100 \%}   &\textbf{100 \%}   &\textbf{100 \%}   &\textbf{100 \%}   &90 \%\\
        \cmidrule(r){1-1}  \cmidrule(r){2-8} \cmidrule(r){9-15}
        WT1 &2 \%   &\textbf{4 \%}   &\textbf{4 \%}   &\textbf{4 \%}   &0 \%   &\textbf{4 \%}   &2 \%
        &94 \%  &94 \%  &90 \%  &96 \%  &92 \%  &\textbf{100 \%}  &94 \%\\
        WT2 &92 \%   &\textbf{94 \%}   &92 \%   &92 \%   &90 \%   &86 \%   &92 \%\\
        WT3 &82 \%   &90 \%   &84 \%   &90 \%   &88 \%   &88 \%   &\textbf{92 \%}\\
        WT4 &42 \%   &56 \%   &\textbf{58 \%}   &\textbf{58 \%}   &52 \%   &38 \%   &50 \%\\
        \cmidrule(r){1-1}  \cmidrule(r){2-8} \cmidrule(r){9-15}
        CC1 &98 \%   &96 \%   &\textbf{100 \%}   &\textbf{100 \%}   &98 \%   &\textbf{100 \%}   &\textbf{100 \%}
        &6 \%  &10 \%  &24 \%  &26 \%  &\textbf{44 \%}  &38 \%  &26 \%\\
        CC2 &92 \%   &96 \%   &\textbf{100 \%}   &96 \%   &96 \%   &88 \%   &84 \%\\
        CC3 &88 \%   &90 \%   &90 \%   &94 \%   &92 \%   &\textbf{100 \%}   &92 \%\\
        CC4 &0 \%   &0 \%   &0 \%   &\textbf{6 \%}   &2 \%   &0 \%   &0 \%\\
        CC5 &\textbf{98 \%}   &96 \%   &96 \%   &\textbf{98 \%}   &96 \%   &\textbf{98 \%}   &84 \%\\
        CCX &42 \%   &72 \%   &\textbf{84 \%}   &80 \%   &80 \%   &80 \%   &62 \%\\
        \cmidrule(r){1-1}  \cmidrule(r){2-8} \cmidrule(r){9-15}
        F16 &\textbf{100 \%}   &98 \%   &\textbf{100 \%}   &\textbf{100 \%}   &\textbf{100 \%}   &96 \%   &76 \%
        &\textbf{98 \%}  &\textbf{98 \%}  &\textbf{98 \%}  &96 \%  &94 \%  &92 \%  &90 \%\\
        \cmidrule(r){1-1}  \cmidrule(r){2-8} \cmidrule(r){9-15}
        SC &\textbf{0 \%}   &\textbf{0 \%}   &\textbf{0 \%}   &\textbf{0 \%}   &\textbf{0 \%}   &\textbf{0 \%}   &\textbf{0 \%}
        &30 \%  &24 \%  &22 \%  &34 \%  &\textbf{42 \%}  &40 \%  &34 \%\\
        \bottomrule
    \end{tabular}}
    \label{tab:resultsrq2}
\end{table*}

\textbf{Results:} Running all of the experiments required approximately \RQonenumberofDays days. We reduced the time to seven days by exploiting the parallelization capabilities of our computing platform.

\Cref{tab:selectionOfThePValue} presents our results. 
Each row of the table refers to one of the requirements.
The table is made of two parts that refer to the assumptions obtained by considering the value ranges $R$ and $R^\prime$.
The cells of the table report the percentage of failure-revealing runs. 
For each requirement and assumption, the value of $p$ that provides the highest percentage of failure-revealing runs is in bold.
If the highest percentage of failure-revealing runs is associated with more than one value of $p$, all these values are in bold. 

\begin{table}[t]
    \centering
     \caption{Minimum, maximum, and variation (difference between the maximum and the minimum) of the percentages of failure-revealing runs from \Cref{tab:selectionOfThePValue} obtained by considering different values for~$p$.}
    \label{tab:variation}
    \footnotesize
    \begin{minipage}{0.45\textwidth}
     \vspace{0pt}
    \begin{tabular}{l r r r r r r }
    \toprule
     &\multicolumn{3}{c}{$\mathbf{R}$}    &\multicolumn{3}{c}{$\mathbf{R}^{\prime}$}\\
     \cmidrule(r){2-4} \cmidrule(r){5-7} 
        & \textbf{max} & \textbf{min} & \textbf{diff} & \textbf{max} & \textbf{min} & \textbf{diff} \\
        \cmidrule(r){1-1} \cmidrule(r){2-4} \cmidrule(r){5-7} 
        AT1 &  $2~\%$ & $0~\%$ & $2~\%$ &  $78~\%$ & $18~\%$ & $60~\%$\\
        AT2 &  $100~\%$ & $98~\%$ & $2~\%$ & $96~\%$ & $88~\%$ & $8~\%$\\
        AT51 & $36~\%$ & $4~\%$ &  $32~\%$\\
        AT52 & $-$ & $-$ &  $-$\\
        AT53 & $98~\%$ & $84~\%$ & $14~\%$\\
        AT54 & $62~\%$ & $4~\%$ & $58~\%$\\
        AT6a & $100~\%$ & $40~\%$ & $60~\%$ & $86~\%$ & $36~\%$ & $50~\%$\\
        AT6b & $96~\%$ & $34~\%$ & $62~\%$ & $56~\%$ & $32~\%$ & $24~\%$\\
        AT6c & $100~\%$ & $54~\%$ & $46~\%$ & $32~\%$ & $6~\%$ & $26~\%$\\
        AT6abc & $98~\%$ & $46~\%$ & $52~\%$ & $32~\%$ & $6~\%$ & $26~\%$\\
        \cmidrule(r){1-1} \cmidrule(r){2-4} \cmidrule(r){5-7}
        AFC27 & $52~\%$ & $38~\%$ & $14~\%$\\
        AFC29 & $-$ & $-$ & $-$\\
        AFC33 & $-$ & $-$ & $-$\\
                \cmidrule(r){1-1} \cmidrule(r){2-4} \cmidrule(r){5-7} 
        NN & $74~\%$ & $66~\%$ & $8~\%$ & $90~\%$ & $78~\%$ & $12~\%$\\
        NNx & $-$ & $-$ & $-$ & $100~\%$ & $90~\%$ & $10~\%$\\
       \bottomrule
               \arrayrulecolor{white}
        \midrule
    \end{tabular}
        \end{minipage}
    \hspace{0.5cm}
        \begin{minipage}{0.45\textwidth}
         \vspace{0pt}
            \begin{tabular}{l r r r r r r }
    \toprule
     &\multicolumn{3}{c}{$\mathbf{R}$}    &\multicolumn{3}{c}{$\mathbf{R}^{\prime}$}\\
     \cmidrule(r){2-4} \cmidrule(r){5-7} 
        & \textbf{max} & \textbf{min} & \textbf{diff} & \textbf{max} & \textbf{min} & \textbf{diff} \\
        \cmidrule(r){1-1} \cmidrule(r){2-4} \cmidrule(r){5-7}
        WT1 & $4~\%$ & $0~\%$ & $4~\%$ & $100~\%$ & $90~\%$ & $10~\%$\\
        WT2 & $94~\%$ & $86~\%$ & $8~\%$\\
        WT3 & $92~\%$ & $82~\%$ & $10~\%$\\
        WT4 & $58~\%$ & $38~\%$ & $20~\%$\\
        \cmidrule(r){1-1} \cmidrule(r){2-4} \cmidrule(r){5-7}
        CC1 & $100~\%$ & $96~\%$ & $4~\%$ & $44~\%$ & $6~\%$ & $38~\%$\\
        CC2 & $100~\%$ & $84~\%$ & $16~\%$\\
        CC3 & $100~\%$ & $88~\%$ & $12~\%$\\
        CC4 & $6~\%$ & $0~\%$ & $6~\%$\\
        CC5 & $98~\%$ & $84~\%$ & $14~\%$\\
        CCX & $84~\%$ & $42~\%$ & $42~\%$\\
        \cmidrule(r){1-1} \cmidrule(r){2-4} \cmidrule(r){5-7}
        F16 & $100~\%$ & $76~\%$ & $24~\%$ & $98~\%$ & $90~\%$ & $8~\%$\\
        \cmidrule(r){1-1} \cmidrule(r){2-4} \cmidrule(r){5-7}
        SC & $-$ & $-$ & $-$ & $42~\%$ & $22~\%$ & $20~\%$\\
        \bottomrule
        \\
        \\
        \\
    \end{tabular}
        \end{minipage}
\end{table}

For $12.8\%$ ($5$ out of $39$) of our assumption-requirement combinations, the value assigned to the parameter $p$ does not influence the percentage of failure-revealing runs.
Specifically, for $2$ models (AT52-R and AFC29-R) \NAME returned a failure revealing test case for $100\%$ of the runs for each value of $p$ and for $3$ models (AFC33-R, NNx-R, and SC-R) \NAME returned a failure revealing test case for $0\%$ of the runs for each value of $p$.
For $87.2\%$ ($34$ out of $39$) of our assumption-requirement combinations, the value assigned to the parameter $p$ influences the percentage of failure-revealing runs.
The assumption-requirement combinations for which the value assigned to $p$ influences the percentages of failure-revealing runs can be further categorized as follows.
For $11.8\%$ ($4$ out of $34$) of these combinations (AT51-R, AT54-R, AT6c-R$^{\prime}$, AT6abc-R$^\prime$), assigning the value $0$ to $p$, i.e., considering only the manual fitness function, led to the highest percentage of failure-revealing runs.
For $2.9\%$ ($1$ out of $34$) of these combinations (WT3-R), assigning the value $1$ to $p$, i.e., considering only the automatic fitness function, lead to the highest percentage of failure-revealing runs.
For $52.9\%$  ($18$ out of $34$) of these combinations (AT53-R, 
AT6b-R, 
AT6c-R, 
AT6abc-R, 
WT1-R, 
WT2-R, 
WT4-R, 
CC2-R, 
CC3-R, 
CC4-R, 
CCX-R, 
AT2-R$^\prime$, 
AT6a-R$^\prime$,
AT6b-R$^\prime$, 
AT6c-R$^\prime$, 
NN-R$^\prime$, 
WT1-R$^\prime$, 
CC1-R$^\prime$, 
SC-R$^\prime$), the highest percentage of failure-revealing runs is obtained by selecting values for $p$ that combine the values returned by the manual and automatic fitness functions, i.e., $0<p<1$.
Therefore, for all of these cases,  using both manual and automatic fitness functions yields better fault detection capabilities.
For the remaining $32.4\%$ ($11$ out of $34$) of the combinations (AT1-R, 
AT2-R, 
AT6a-R, 
AFC27-R, 
NN-R, 
CC1-R, 
CC5-R, 
F16-R, 
AT1-R$^\prime$, 
NNx-R$^\prime$, 
F16-R$^\prime$), the values of $p$ that lead to the highest percentage of failure-revealing runs include either the value $0$ or $1$.

For the $87.2\%$ ($34$ out of $39$) of our assumption-requirement combinations, for which the value assigned to the parameter $p$ influences the percentage of failure-revealing runs, 
\Cref{tab:variation} reports the variation in the percentage of failure-revealing runs detected by considering different values for $p$.
Specifically, it reports the minimum and maximum  percentages of failure-revealing runs obtained by considering different values for~$p$ and their difference (variation).
The variation in the percentage of failure-revealing runs across the assumption-requirement combinations we considered ranges from \minvariation to
\maxvariation\ runs ($\mathit{avg}=\avgvariation$, $\mathit{sd}=\stdvariation$).

The optimal value for the parameter $p$ for the \NAME fitness function changes across the different assumption-requirement combinations.
For each combination, the values of $p$ that provide the highest percentage of failure-revealing runs are in bold.
Since the behavior of \NAMESIMULINK depends on the value assigned to the parameter $p$, for \NAMESIMULINK, 
we considered two configurations: 
$\NAMESIMULINK_{avg}$ and $\NAMESIMULINK_{best}$.
The configuration $\NAMESIMULINK_{avg}$ sets $0.5$ as the value for the parameter $p$ since $0.5$ is the average across the different assumption-requirement combinations of the values of $p$ with the highest failure-revealing capabilities.
For combinations that contain multiple values of $p$ leading to the highest failure-revealing capabilities, we considered their average for selecting the parameter $p$.
For example, for AT1, the values for $p$ leading to the highest failure-revealing capability are $0$, $0.2$, $0.4$, and $0.5$.
Therefore, we considered the value $0.275$ for the computation of the average value of the parameter $p$.
The configuration $\NAMESIMULINK_{best}$ selects for each assumption-requirement combination the best value for the parameter $p$ (when more than one value to $p$ leads to the highest effectiveness, one of these values was randomly selected).\\
The Wilcoxon signed-rank test~\cite{woolson2007wilcoxon} (with significance level $\alpha=0.05$) confirms that $\NAMESIMULINK_{best}$ generates a different number of failure-revealing runs when compared to every value of $p$.

\begin{Answer}[RQ2 - \NAME Fitness Selection]
The selection of the \NAME fitness function influences the percentages of failure-revealing runs returned by \NAME for $87.2\%$ ($34$ out of $39$) of the assumption-requirement combinations we considered.
For these assumption-requirement combinations, the variation in the percentage of failure-revealing runs
ranges from \minvariation to
\maxvariation\ runs ($\mathit{avg}=\avgvariation$, $\mathit{sd}=\stdvariation$).

For each assumption-requirement combination, we identified the optimal value for the parameter $p$.
We identified two configurations of \NAME to be used in the rest of our experiments: $\NAMESIMULINK_{avg}$ uses the average,  across the different assumption-requirement combinations, of the optimal values of $p$, $\NAMESIMULINK_{best}$ uses (one of) the best values of $p$ for each assumption-requirement combination.
\end{Answer}

\subsection{Effectiveness --- RQ3}
\label{sec:rq1}
We compare the effectiveness of  \NAMEAUTOMATIC, \NAMEMANUAL, and \NAMESIMULINK in generating failure-revealing test cases.

\textbf{Software Configuration}:
For \NAMEAUTOMATIC, \NAMEMANUAL, and \NAMESIMULINK, we used the search algorithm and the maximum number of iterations considered in RQ2. 
For \NAMEMANUAL and \NAMESIMULINK, we considered the manual fitness functions identified in RQ1.
For \NAMESIMULINK, we considered the two configurations ($\NAMESIMULINK_{avg}$ and $\NAMESIMULINK_{best}$) from RQ1.
Recall that \NAMEAUTOMATIC and \NAMEMANUAL are specific instances of \NAMESIMULINK where the parameter $p$ is respectively set to the values $1$ and $0$. 

\textbf{Methodology}:  
For  each of the \numrequirementassumption assumption-requirement combinations from \Cref{tab:assumptions}, 
we analyzed the results obtained by using \NAMEAUTOMATIC, \NAMEMANUAL,  $\NAMESIMULINK_{avg}$ and $\NAMESIMULINK_{best}$ reported in \Cref{tab:selectionOfThePValue}.

\textbf{Results:} For each tool and assumption-requirement combination, Table~\ref{tab:comparison} presents the percentage of failure-revealing  runs (over \nruns runs) for \NAMEAUTOMATIC, \NAMEMANUAL,  $\NAMESIMULINK_{avg}$ and $\NAMESIMULINK_{best}$.
The table reports the results obtained by considering the assumptions with the value range $R$ and $R^\prime$.
For example, for the AT1 requirement, \NAMEAUTOMATIC returned a failure-revealing test case for $0 \%$ and $18 \%$ of the runs for the assumptions obtained from the  value ranges $R$ and $R^\prime$ respectively.

For $\approx74\%$ of the  combinations (\RQonenumathenawins out of \numrequirementassumption),
the percentage of failure-revealing runs of $\NAMESIMULINK_{avg}$ is greater than or equal to that of \NAMEAUTOMATIC and \NAMEMANUAL. 
For these combinations, $\NAMESIMULINK_{avg}$ has to be preferred since, in the worst case, it works as the best among \NAMEAUTOMATIC and \NAMEMANUAL.
On average $\NAMESIMULINK_{avg}$ reveals \RQonewinsathenavsstaliro (\textit{min}=\RQonewinsathenavsstaliromin, \textit{max}=\RQonewinsathenavsstaliromax, \textit{StdDev}=\RQonewinsathenavsstalirostd) more failures when compared with \NAMEAUTOMATIC, and 
\RQonewinsathenavsmanual (\textit{min}=\RQonewinsathenavsmanualmin, \textit{max}=\RQonewinsathenavsmanualmax,  \textit{StdDev}=\RQonewinsathenavsmanualstd) when compared with \NAMEMANUAL.

Remarkably, unlike \NAMEAUTOMATIC and \NAMEMANUAL, $\NAMESIMULINK_{avg}$  generated failure-revealing test cases for CC4, i.e.,  $\NAMESIMULINK_{avg}$ detected failures other tools could not find.

\begin{table*}[t]
    \centering
    \footnotesize
    \caption{Percentage of failure-revealing runs of 
    \NAMEMANUAL (\textbf{SM}),
    \NAMEAUTOMATIC (\textbf{SA})
$\NAMESIMULINK_{\mathbf{avg}}$ ($\mathbf{S_{\mathbf{avg}}}$),
$\NAMESIMULINK_{\mathbf{best}}$ ($\mathbf{S_{best}}$) and assumption-requirement combination of Table~\ref{tab:assumptions}.}
    \label{tab:comparison}
    \begin{tabular}{l r r r r r r r r  }
    \toprule
    &\multicolumn{4}{c}{$\mathbf{R}$}    &\multicolumn{4}{c}{$\mathbf{R}^{\prime}$}\\
     \cmidrule(r){2-5} \cmidrule(r){6-9} 
        \textbf{RID} & 
  \multicolumn{1}{c}{\textbf{SM}} & 
    \multicolumn{1}{c}{\textbf{\textbf{SA}}} &
        \multicolumn{1}{c}{\textbf{$\mathbf{S_{\mathbf{avg}}}$}}
        & \multicolumn{1}{c}{$\mathbf{S_{best}}$} &  
        \multicolumn{1}{c}{\textbf{SM}} & \multicolumn{1}{c}{\textbf{\textbf{SA}}}  & \multicolumn{1}{c}{\textbf{$\mathbf{S_{\mathbf{avg}}}$}}
        & \multicolumn{1}{c}{$\mathbf{S_{best}}$}  
        \\
        \cmidrule(r){1-1}  \cmidrule(r){2-5} \cmidrule(r){6-9} 
        AT1     & 2 \%  & 0 \%  & 2 \% & 2 \% & 78 \% & 18 \% & 78 \% & 78 \%\\
        AT2     & 100 \% & 100 \% & 100 \% & 100 \% & 88 \% & 92 \% & 92 \% & 96 \% \\
        AT51    & 36 \% & 4 \% & 12 \% & 36 \% \\
        AT52    & 100 \% & 100 \% & 100 \% & 100 \% \\
        AT53    & 94 \% & 84 \% & 94 \% & 98 \% \\
        AT54    & 62 \% & 4 \% & 50 \% & 62 \% \\
        AT6a    & 40 \% & 100 \% & 96 \% & 100 \% & 36 \% & 58 \% & 86 \% & 86 \% \\
        AT6b    & 34 \% & 88 \% & 92 \% & 96 \% & 32 \% & 38 \% & 56 \% & 56 \%\\
        AT6c    & 54 \% & 90 \% & 90 \% & 100 \% & 32 \% & 6 \% & 18 \% & 32 \%\\
        AT6abc & 46 \% & 96 \% & 98 \% & 98 \% & 32 \% & 6 \% & 18 \% & 32 \%\\
              \cmidrule(r){1-1}  \cmidrule(r){2-5} \cmidrule(r){6-9} 
        AFC27 & 52 \% & 38 \% & 48 \% & 52 \% \\
        AFC29  & 100 \%  &100 \% & 100 \% & 100 \% \\
        AFC33  & 0 \% & 0 \% & 0 \% & 0 \%\\
              \cmidrule(r){1-1}  \cmidrule(r){2-5} \cmidrule(r){6-9} 
        NN  & 68 \% & 74 \% & 68 \% & 74 \% & 86 \% & 78 \% & 82 \% & 90 \% \\
        NNx  &0 \% & 0 \% &0 \% & 0 \% & 100 \% & 90 \% & 100 \% & 100 \%\\
             \cmidrule(r){1-1}  \cmidrule(r){2-5} \cmidrule(r){6-9} 
        WT1  & 2 \% & 2 \% & 4 \% & 4 \% & 94 \% & 94 \% & 96 \% & 100 \%\\
        WT2  & 92 \% & 92 \% & 92 \% & 94 \% \\
        WT3  & 82 \% & 92 \% & 90 \% & 92 \% \\
        WT4  & 42 \% & 50 \% & 58 \% & 58 \%\\
             \cmidrule(r){1-1}  \cmidrule(r){2-5} \cmidrule(r){6-9} 
        CC1     & 98 \% & 100 \% & 100 \% & 100 \% & 6 \% & 26 \% & 26 \% & 44 \%\\
        CC2     & 92 \% & 84 \% & 96 \% & 100 \%\\
        CC3     & 88 \% & 92 \% & 94 \% & 100 \% \\
        CC4     & 0 \% & 0 \% & 6 \% & 6 \%\\
        CC5     & 98 \% & 84 \% & 98 \% & 98 \%\\
        CCX     & 42 \% & 62 \% & 80 \% & 84 \%\\
              \cmidrule(r){1-1}  \cmidrule(r){2-5} \cmidrule(r){6-9} 
        F16     & 100 \% & 76 \% & 100 \% & 100 \% & 98 \% & 90 \% & 96 \% & 98 \% \\
               \cmidrule(r){1-1}  \cmidrule(r){2-5} \cmidrule(r){6-9} 
        SC      &0 \% & 0 \% & 0 \% & 0 \% & 30 \% & 34 \% & 34 \% & 42 \% \\
    \bottomrule
    \end{tabular}
\end{table*}

For only $\approx26\%$ of the assumption-requirement combinations (\RQonenumathenaloses out of \numrequirementassumption), $\NAMESIMULINK_{avg}$ generated less  failure-revealing runs than (at least one of) the baselines. 
For these cases, the penalty of $\NAMESIMULINK_{avg}$ is negligible — 
the percentage of  failure-revealing runs is only 
$24\%$ (for AT51), 
$12\%$ (for AT54), 
$4\%$ (for AT6a), 
$14\%$ (for AT6c), 
$14\%$ (for AT6abc), 
$4\%$ (for AFC27),
$6\%$ and $4\%$ (for NN),
$2\%$ (for WT3),
and $2\%$ (for F16)
lower than the best baseline framework.
The decrement of $\NAMESIMULINK_{avg}$ in the percentage of failure-revealing runs is on
average \RQonelosesathenavsstaliro (\textit{min}=\RQonelosesathenavsstaliromin, \textit{max}=\RQonelosesathenavsstaliromax, \textit{StdDev}=\RQonelosesathenavsstalirostd) when compared with \NAMEAUTOMATIC and 
\RQonelosesathenavsmanual (\textit{min}=\RQonelosesathenavsmanualmin, \textit{max}=\RQonelosesathenavsmanualmax,  \textit{StdDev}=\RQonelosesathenavsmanualstd) when compared with \NAMEMANUAL.

Considering all the \numrequirementassumption assumption-requirement combinations,  the percentage of  failure-revealing runs of $\NAMESIMULINK_{avg}$  is on average \RQoneathenavsstaliro  (\textit{min}=\RQoneathenavsstaliromin,
\textit{max}=\RQoneathenavsstaliromax,
\textit{StdDev}=\RQoneathenavsstalirostd)  and 
\RQoneathenavsmanual 
(\textit{min}=\RQoneathenavsmanualmin,
\textit{max}=\RQoneathenavsmanualmax, 
\textit{StdDev}=\RQoneathenavsmanualstd) higher than \NAMEAUTOMATIC and \NAMEMANUAL. 

Since $\NAMESIMULINK_{best}$ always selects the best value for the parameter $p$, for all the  combinations (\numrequirementassumption out of \numrequirementassumption),
the percentage of failure-revealing runs of $\NAMESIMULINK_{best}$ is higher than or equal to the one of \NAMEAUTOMATIC and \NAMEMANUAL. 
The increment $\NAMESIMULINK_{best}$ provides on the percentage of failure-revealing runs is on average \RQonewinsathenabestvsstaliro (\textit{min}=\RQonewinsathenabestvsstaliromin, \textit{max}=\RQonewinsathenabestvsstaliromax, \textit{StdDev}=\RQonewinsathenabestvsstalirostd) when compared with \NAMEAUTOMATIC, and 
\RQonewinsathenabestvsmanual (\textit{min}=\RQonewinsathenabestvsmanualmin, \textit{max}=\RQonewinsathenabestvsmanualmax,  \textit{StdDev}=\RQonewinsathenabestvsmanualstd) when compared with \NAMEMANUAL.

Similarly to $\NAMESIMULINK_{avg}$,  $\NAMESIMULINK_{best}$  generated failure-revealing test cases for CC4 that \NAMEAUTOMATIC and \NAMEMANUAL could not find.

The Wilcoxon signed-rank test~\cite{woolson2007wilcoxon} (with significance level $\alpha=0.05$) confirms that $\NAMESIMULINK_{best}$ and $\NAMESIMULINK_{avg}$ generate more failure-revealing runs than \NAMEAUTOMATIC 
and \NAMEMANUAL. 
The Vargha-Delaney effect size test~\cite{varghadelaney} confirms that there is evidence of stochastic superiority of $\NAMESIMULINK_{best}$ and $\NAMESIMULINK_{avg}$ compared to \NAMEAUTOMATIC and \NAMEMANUAL.
We conclude that \NAMESIMULINK can generate more failure-revealing runs than \NAMEAUTOMATIC and \NAMEMANUAL.

\newpage
\begin{Answer}[RQ3 - Effectiveness]
\NAMESIMULINK is preferable over the baseline tools for between $\approx74\%$ and $100\%$ of our assumption-requirement combinations depending on the value selected for the parameter $p$.
When the \NAME fitness equally prioritizes the manual and automatic fitness functions,
\NAMESIMULINK  generated on
average \RQoneathenavsstaliro  (\textit{min}=\RQoneathenavsstaliromin, 
\textit{max}=\RQoneathenavsstaliromax, 
\textit{StdDev}=\RQoneathenavsstalirostd)  and 
\RQoneathenavsmanual 
(\textit{min}=\RQoneathenavsmanualmin, 
\textit{max}=\RQoneathenavsmanualmax, 
\textit{StdDev}=\RQoneathenavsmanualstd) more failure-revealing runs than \NAMEAUTOMATIC and \NAMEMANUAL. 
When the \NAME prioritization of the manual and automatic fitness functions is tailored to the specific models,
\NAMESIMULINK  generated on
average  \RQonewinsathenabestvsstaliro  (\textit{min}=\RQonewinsathenabestvsstaliromin,
\textit{max}=\RQonewinsathenabestvsstaliromax,
\textit{StdDev}=\RQonewinsathenabestvsstalirostd)  and 
\RQonewinsathenabestvsmanual 
(\textit{min}=\RQonewinsathenabestvsmanualmin,
\textit{max}=\RQonewinsathenabestvsmanualmax, 
\textit{StdDev}=\RQonewinsathenabestvsmanualstd) more failure-revealing runs than \NAMEAUTOMATIC and \NAMEMANUAL. 

For one combination, unlike \NAMEAUTOMATIC and \NAMEMANUAL, \NAMESIMULINK generated failure-revealing test cases.
\end{Answer}

\subsection{Efficiency --- RQ4}
\label{sec:rq2}
We compare the efficiency of \NAMEAUTOMATIC, \NAMEMANUAL, and \NAMESIMULINK in generating failure-revealing test cases to confirm that \NAMESIMULINK is not less performant than  \NAMEAUTOMATIC and \NAMEMANUAL.

\textbf{Software Configuration}: Since our goal is to confirm that \NAMESIMULINK does not perform worse than \NAMEAUTOMATIC and \NAMEMANUAL, we consider instances of \NAMESIMULINK obtained by assigning $p$ to $0$, $0.2$, $0.4$, $0.5$, $0.6$, $0.8$, and $1$ and $\NAMESIMULINK_{best}$. 
Recall that instances where $p$ is assigned to the values $0$, $0.5$, and $1$ respectively correspond to \NAMEMANUAL, $\NAMESIMULINK_{avg}$, and \NAMEAUTOMATIC.

\textbf{Methodology}:  
We analyze the results of RQ2 and consider only the failure-revealing runs.
Since for each iteration, the difference in the execution time of \NAMEAUTOMATIC, \NAMEMANUAL, and \NAMESIMULINK is negligible, we use the number of iterations as the metric to compare the efficiency of the considered tools.
For each value of the parameter $p$, we analyze the failure-revealing runs and extract the number of iterations required to generate the failure-revealing test cases.

\begin{table*}
    \centering
    \caption{Number of failure-revealing runs (\# Samples), median (Med Iter), and average (Avg Iter) number of iterations per run for each value of $p$.}
    \footnotesize
    \[
    \begin{array}{l r r r r r r r r}
        \toprule
        \boldsymbol{p}           &\boldsymbol{0}      &\boldsymbol{0.2}    &\boldsymbol{0.4}    &\boldsymbol{0.5}    &\boldsymbol{0.6}    &\boldsymbol{0.8}    &\boldsymbol{1}      &\boldsymbol{p_{best}} \\
        \cmidrule(r){1-1}   \cmidrule(r){2-8}   \cmidrule(r){9-9}
        \text{\# Samples}   &1118   &1205   &1259   &1275   &1268   &1230   &1121   &1354\\
        \text{Med Iter}     &90     &96     &95     &96     &103    &97     &93     &102 \\
        \text{Avg Iter}     &107.8  &109.3  &108.6  &108.3  &111.2  &109.6  &107.5  &113.5 \\
        \bottomrule
    \end{array}
    \]
    \label{tab:NIter}
\end{table*}

\begin{figure}[t]
    \includegraphics[width=0.9\columnwidth]{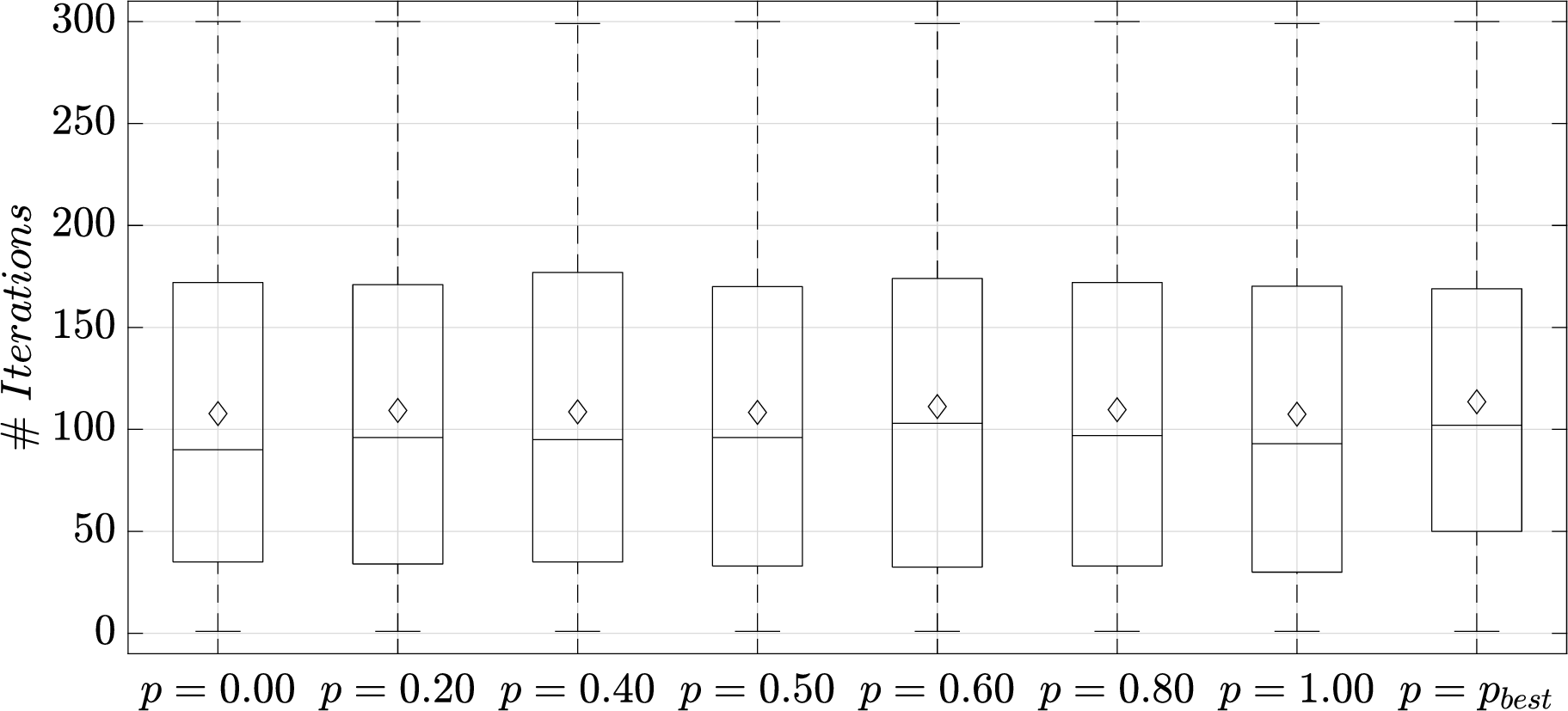}
    \caption{Number of iterations of \NAMEAUTOMATIC ($p=1$), \NAMEMANUAL ($p=0$), and \NAME ($0 \leq p \leq 1$).
    Diamonds depict the average.
    }
    \label{fig:rq2results}
    \Description[Boxplots showing the number of iterations for different values of $p$.]{Boxplots showing the number of iterations for different values of $p$. The plot has a separate column for each value of $p$ and a last column for $p_{best}$. The plot shows that for all the columns, the median number of iterations is around 100, with a slightly higher average number of iterations.}
\end{figure}

\textbf{Results:} 
The box plots in \Cref{fig:rq2results} show the distribution in the number of iterations required by \NAMEAUTOMATIC ($p=1$), \NAMEMANUAL ($p=0$), \NAMESIMULINK ($0 \leq p \leq 1$) and $\NAMESIMULINK_{best}$ ($p=p_{best}$). 
The Wilcoxon rank sum test~\cite{mcdonald2009handbook} (Significance Level $\alpha=0.05$) confirms that there is no statistical difference between the median number of iterations $\NAMESIMULINK_{avg}$, \NAMEAUTOMATIC, and \NAMEMANUAL
require to generate failure-revealing test cases.
It also confirms that the median number of iterations $\NAMESIMULINK_{best}$ requires to generate failure-revealing test cases is higher than the ones needed by \NAMEAUTOMATIC 
and \NAMEMANUAL.
However, as listed in \Cref{tab:NIter}, the difference in the number of iterations between $\NAMESIMULINK_{best}$ and any other value of $p$ is at most $12$ iterations for the median and $6.1$ iterations for the average.
The most computation-intensive model (F16) requires only $3s$ per iteration.
Therefore, $12$ additional iterations correspond to less than a minute of running time. 
This overhead is negligible, considering that the development of CPS models usually takes months or years.
Note that our results are not in contradiction with the results from RQ3 since (a)~efficiency and effectiveness are two dimensions of the problem (being more effective in finding failure-revealing test cases within a given time budget does not require being more efficient), and (b)~to generate the box plots we only considered the failure-revealing runs.
Therefore, the number of runs considered for generating the box plots of $\NAMESIMULINK_{avg}$ and $\NAMESIMULINK_{best}$ is higher since they are more effective than \NAMEAUTOMATIC and \NAMEMANUAL in generating failure-revealing runs (see  \text{\# Samples} in \Cref{tab:NIter}). 
For example,  unlike \NAMEAUTOMATIC and \NAMEMANUAL, $\NAMESIMULINK_{avg}$ and $\NAMESIMULINK_{best}$  generate failure-revealing test cases for CC4 requiring on average $260$ iterations each. 
We considered these failure-revealing runs in generating the box plots reported in \Cref{fig:rq2results}.
This decision penalizes $\NAMESIMULINK_{avg}$ and $\NAMESIMULINK_{best}$ over \NAMEAUTOMATIC and \NAMEMANUAL in assessing efficiency.

\begin{Answer}[RQ4 - Efficiency]
For our assumption-requirement combinations, the difference between the number of iterations required by \NAMEAUTOMATIC, \NAMEMANUAL, and \NAMESIMULINK to generate the failure-revealing test cases is not statistically relevant or negligible depending on the value assigned to the parameter $p$.
\end{Answer}

\subsection{Usefulness --- RQ5}
\label{sec:rq3}
To assess the usefulness of \NAMESIMULINK, we evaluate its applicability in two case studies from the automotive and medical domains. 

\subsubsection{Automotive case study}
\label{sub:RQ5auto}
\hfill\\

\textbf{Benchmark:}
Our automotive case study is the \simulink model of a hybrid-electric vehicle (HEV)  developed for the EcoCAR Mobility Challenge~\cite{EcoCAR}, a competition sponsored by the U.S. Department of Energy~\cite{DOE}, General Motors~\cite{GM}, and MathWorks~\cite{MathWorks}.
The HEV motor converts electrical energy into mechanical energy. 
A software controller regulates the behavior of the motor. 

The HEV model includes 4775 blocks. 
It consists of multiple subsystems built using Add-On components of \simulink, including Simscape~\cite{simscape}, Simscape Electrical~\cite{simscapeElectrical}, and Simscape Driveline~\cite{Driveline}.
The controller is modeled with \simulink Stateflow~\cite{Stateflow}.
The controller input is the speed demand, i.e., the required speed in Kph (kilometer per hour), and its output is the vehicle speed.
The HEV model contains three input exemplars with the speed demand for three urban driving scenarios.

\textbf{Methodology:}
To generate realistic driving scenarios, we considered one of the three urban driving scenario examples and slightly varied the speed demand. 
Specifically, we added the input signal \texttt{delta\_i\_speed} to the HEV model that represents the variation applied to the speed demand of the urban driving scenario we considered.
We set the value $400$s for the simulation time since it is the simulation time provided for the HEV model. 

To use \NAMESIMULINK, we first needed to design assumptions for the \texttt{delta\_i\_speed} input signal. 
We set $[0,4]$Kph as the value range for the assumption since it is a sufficiently small range for the variation of the speed demand.
We considered five control points to ensure speed variations occur every $100$s. 
We set the \texttt{pchip} interpolation function~\cite{PCHIP} since it generates smooth and continuous signals for the variations of the speed demand.

The requirement we considered specifies that the difference between the desired speed and the speed of the vehicle after $0.7$s shall be lower than a threshold value. 
We declared an output signal $\texttt{delta\_o\_speed}$ representing the difference between the vehicle speed and the desired demand $0.7$s before. 
Then, we expressed the requirement in STL as
$\LTLg_{[0,400]}(\texttt{delta\_o\_speed}<\texttt{threshold})$.
The temporal operator $\LTLg_{[0,400]}$ requires the value of $\texttt{delta\_o\_speed}$ to be lower than the threshold value within the interval $[0,400]$s.
We set $3$Kph as the threshold value since, for the urban driving scenario we considered,  $\texttt{delta\_o\_speed}$ is always lower than this value. 

We prioritized inputs that generated considerable changes for the speed demand since we believed that these inputs were likely to lead to a violation of our requirement.
Therefore, we define a manual fitness function that returns a value that decreases as the average distance between consecutive control points increases. 
We used the implementation for the method \texttt{athenaFitness} from Listing~\ref{listingMatlab} and set $0.5$ as the value for $p$ to equally prioritize the manual and the automatic fitness functions.
We set the value to $300$ for the number of iterations. Then, we ran \NAMESIMULINK once.
As expected,  \NAMESIMULINK could not generate any failure-revealing test case.
We injected a representative fault in the model: we changed the threshold that makes the car switch from the \texttt{Cruise\_mode} to the \texttt{Accelerate\_mode} by $50\%$.
We then ran \NAMESIMULINK again and checked whether it could generate failure-revealing test cases for the faulty model.

\begin{figure*}[t]
    \centering
    \begin{subfigure}[b]{0.49\columnwidth}
        \centering
        \includegraphics[width = 0.9\textwidth]{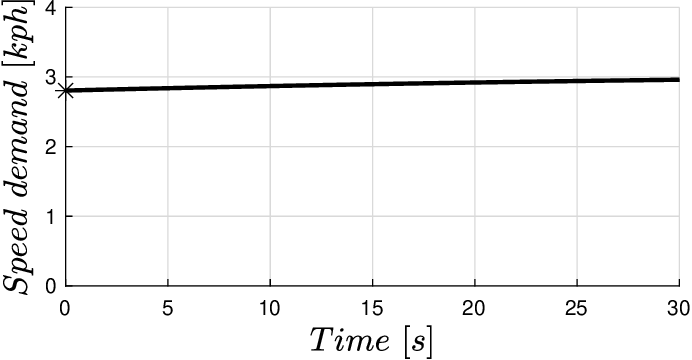}
        \caption{Failure-revealing speed demand variation for the HEV model.}
        \label{fig:InputHEV}
        \Description[Plot of the Speed demand variation.]{Plot of the Speed demand variation for the first 30 seconds of simulation. The plot shows an almost constant signal at 3 kph.}
    \end{subfigure}
    \hfill
    \begin{subfigure}[b]{0.49\columnwidth}
        \centering
        \includegraphics[width = 0.9\textwidth]{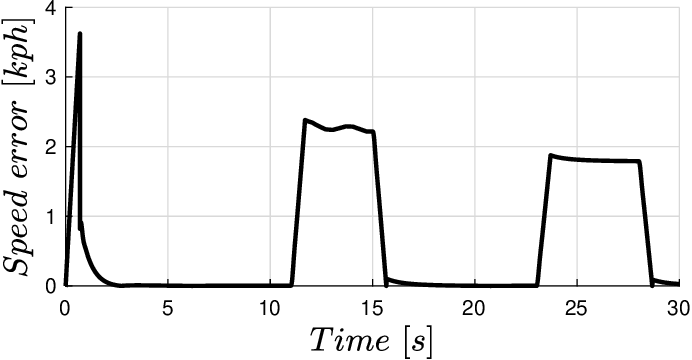}
        \caption{Output signals for the HEV  model for the input of \Cref{fig:InputHEV}.}
        \label{fig:OutputHEV}
        \Description[Plot of the Speed error.]{Plot of the Speed error for the first 30 seconds of simulation. The plot shows a peak in the signal at the beginning of the simulation, that exceeds 3 kph.}
    \end{subfigure}
    
    \caption{Example input and output for the Hybrid Electric Vehicle (HEV) \simulink model for the interval $[0,30]s$.}
    \label{fig:HybridSignals}
\end{figure*}

\textbf{Results:}
\NAMESIMULINK generated a failure-revealing test in $12$ iterations requiring $199$s ($\approx 3$min).
\Cref{fig:HybridSignals} shows a portion of the failure-revealing input and output signals. 
The failure-revealing test case shows that if the speed demand changes substantially within the first $5$ seconds of the simulation, the controller cannot keep the difference between the vehicle speed and the desired demand $0.7$s before within the desired threshold ($3$Kph).
The cause of the problem was the change we made to the model: increasing the threshold that makes the car switch from the \texttt{Cruise\_mode} to the \texttt{Accelerate\_mode} by $50\%$ causes a problem when there is a variation of the speed at the beginning of the simulation.

\subsubsection{Medical case study}
\label{sub:RQ5med}
\hfill\\

\textbf{Benchmark:}
Our medical case study is a mechanical ventilator~\cite{Ventilator} developed by MathWorks.
The mechanical ventilator assists in the breathing of patients.
The model we consider includes 241 blocks and consists of a real-time controller and a system model.
The real-time controller is modeled in \simulink Stateflow~\cite{Stateflow}. 
The system model includes the lungs and trachea of the patient.
This example provides a starting point for designers working on ventilators, showing how to interface a real-time controller and a system model.

\textbf{Methodology:}
We considered muscle pressure, body temperature, and room temperature as external inputs.
The muscle pressure signal estimates the total inspiratory muscle pressure of a patient~\cite{Pleil_2021}.
The body temperature and the room temperature are, respectively, the temperature of the body of the patient and the room.

The assumptions we considered for our input signals are as follows. 
We set $[0,0.05]$bar, $[35,39]^{\circ}$C, and $[18,25]^{\circ}$C as value range for the muscle pressure, body temperature, and room temperature external inputs.
The maximum muscle pressure is the expiratory pressure for an average healthy adult woman~\cite{MusclePressure}; the body and room temperature are assigned to reasonable value ranges.
We considered two control points for the muscle pressure and used a step function as an interpolation function to simulate the sudden changes in the inhaling activity of the patient.
We used two constant values for the body and room temperatures since we assumed these values do not change over the simulation time ($30$s).  

The requirement specifies that the mechanical ventilator should ensure that the internal pressure of the patient's respiratory system is below a threshold value and that the limits of the lung volume are not exceeded.
We express this requirement in STL as
$\LTLg_{[0,30]}(\texttt{pressure}\leq\texttt{threshold}\land\texttt{min\_volume}\leq \texttt{volume}\leq\texttt{max\_volume})$.
The temporal operator $\LTLg_{[0,30]}$ requires the value of the pressure ($\texttt{pressure}$) and the lung volume ($\texttt{volume}$) to be within the threshold values during the interval $[0,30]$s.
We set $30cm_{H_{2}O}$ as the threshold value for the pressure since it is the recommended threshold for the Plateau Pressure (PP) to avoid lung injury in patients connected to a mechanical ventilator \cite{PlateauPressure}.
We set $0$L and $6$L as minimum and maximum threshold values for the lung volume since $6$L is the average lung capacity for an adult \cite{TotalLungCapacity}.

Our manual fitness function prioritizes inputs with high initial and low final pressure as well as high body temperature.
We select this fitness function to test if the ventilator can handle sudden changes in the state of the patient, i.e., suddenly stopping the inhaling activity, and can ensure its safety in critical conditions, like high body temperature and irregular breathing patterns.

\begin{figure*}[t]
    \centering
    \begin{subfigure}[b]{0.49\columnwidth}
        \centering
        \includegraphics[width = 0.9\textwidth]{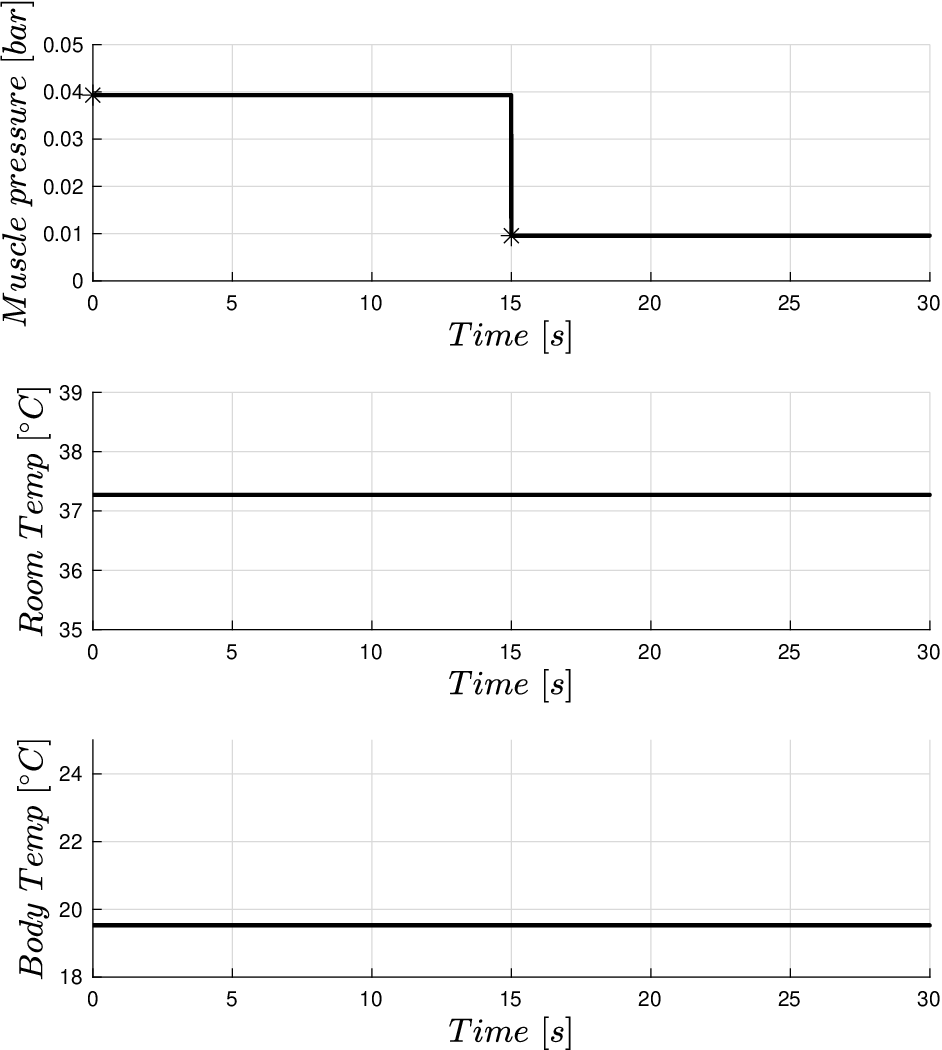}
        \caption{Failure-revealing input signals for the MV model.\phantom{RandomWord}}
        \label{fig:InputMV}
        \Description[Plot of the inputs of the Mechanical Ventilator.]{Plot of the inputs of the Mechanical Ventilator. The plots show a constant body temperature of $37$ degrees, a room temperature of $19$ degrees, and a step muscular pressure going from $0.04~bar$ to $0.01~bar$ at $15$ seconds.}
    \end{subfigure}
    \hfill
    \begin{subfigure}[b]{0.49\columnwidth}
        \centering
        \includegraphics[width = 0.9\textwidth]{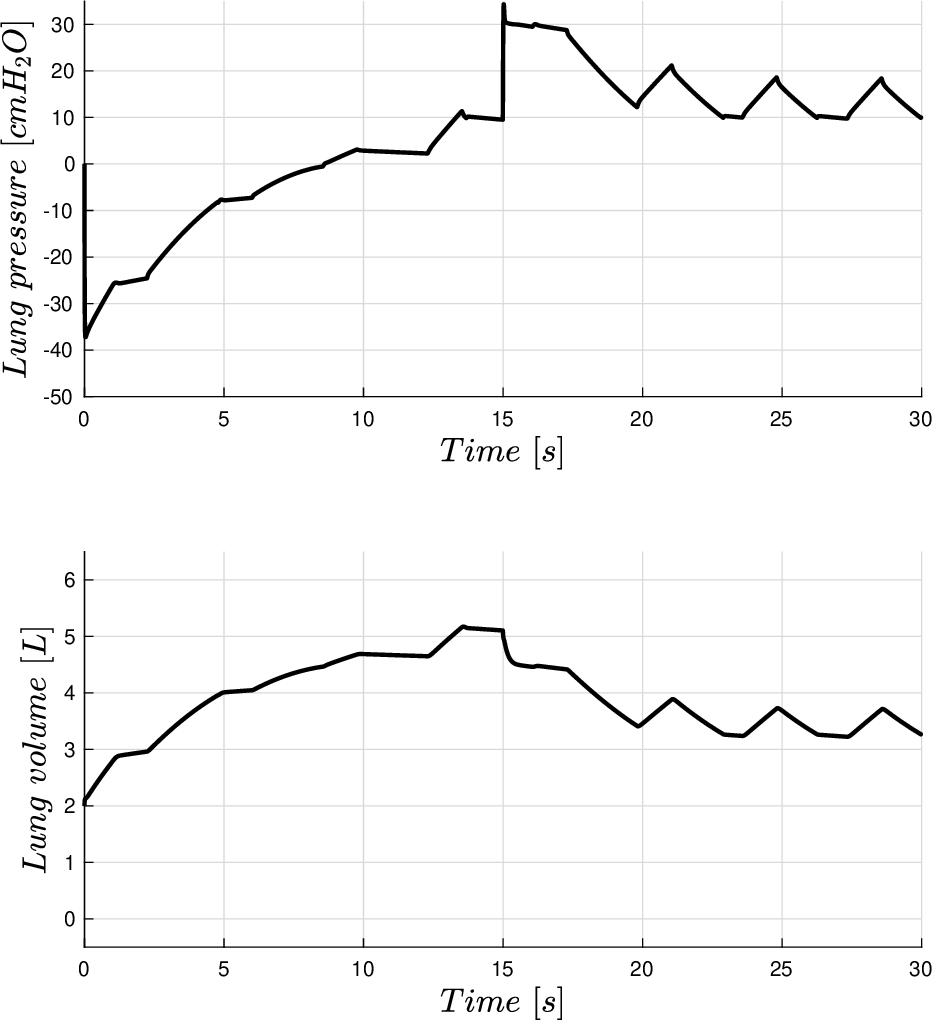}
        \caption{Output signals for the MV  model for the input of \Cref{fig:InputMV}.}
        \label{fig:OutputMV}
        \Description[Plot of the outputs of the Mechanical Ventilator.]{Plot of the outputs of the Mechanical Ventilator. The plots show a sudden increase in lung pressure at $15$ seconds.}
    \end{subfigure}
    
    \caption{Example input and output for the Mechanical Ventilator (MV) \simulink model.}
    \label{fig:MechVentSignals}
\end{figure*}

\textbf{Results:}
\NAMESIMULINK generated a failure-revealing test in $8$ iterations requiring $452$s ($\approx 7.5$min).
\Cref{fig:MechVentSignals} shows a portion of the failure-revealing input and output signals.
The figure shows that when the voluntary breathing of the patient is abruptly reduced (see time instant $t=15s$), the mechanical ventilator cannot guarantee that the internal pressure of the respiratory system of the patient is below a threshold value ($30cm_{H_{2}O}$); it reaches $34.3cm_{H_{2}O}$ at time instant $15.0$.
Therefore, although the ventilator relies on the patient's inhalation and exhalation cycle to control the pressure, a sudden change can lead to a violation of the pressure threshold, which could damage the respiratory tract of the patient.
We remark that although MathWorks experts developed the model, they did not ensure the correctness of their model — the goal of the model is to provide a starting point for designers working on ventilators, showing how to interface a real-time controller and a system model.

\begin{Answer}[RQ5 - Usefulness]
\NAMESIMULINK was able to compute a failure-revealing test case for two case studies from the automotive and medical domains. 
 \NAMESIMULINK returned these test cases respectively after $12$ and $8$ iterations that required approximately three and seven minutes of computation.
\end{Answer}



\section{Discussion and Threats to Validity}
\label{sec:discussion}
For the benchmark models and requirements we considered, our results show that \NAMESIMULINK is more effective in detecting failure-revealing test cases than the baseline frameworks with no significant runtime performance overhead.
Additionally, \NAMESIMULINK was able to generate a failure-revealing test case for CC4 that the baselines could not find.
Finally, \NAMESIMULINK was able to generate a failure-revealing test case for our case studies.
Based on these results, a possible workflow for the use of \NAME is as follows: 
Engineers should initially use SBST frameworks based on automated fitness functions since they do not require manual effort to be defined and may already return failure-revealing test cases. If no failure-revealing test case is detected, engineers should use SBST frameworks based on manually defined fitness functions since they guide the search toward portions of the input domain more likely to contain failures.
Finally, engineers should use SBST frameworks that combine automatically generated and manually defined fitness functions and can detect failures other frameworks cannot find.
Note that this workflow is only one of the possible workflows for using \NAME, which is proposed based on the results of our empirical investigation. Future experiments and empirical analysis may confirm the benefits of this workflow in practice or suggest new workflow usages for \NAME.

In the following section, we reflect on the novelty of the proposed solution, the definition and use of the manual fitness functions, the applicability of the automatic fitness functions, and \NAME in general.
Finally, we discuss the threats to the validity of our results.

\textbf{Novelty}: Developing techniques that enable the generation of failure-revealing test cases effectively and efficiently is
a widely known software engineering problem~\cite{papadakis2019mutation,5210118}.
No existing solution enables engineers to combine automatically generated and manually defined fitness functions.  Therefore, \NAME is a significant technical contribution: it improves the research literature and practice by introducing a novel and new solution and extensively evaluates the solution on existing benchmarks from the literature and two representative case studies.

\textbf{Manual fitness function definition and use}: To define the manual fitness function, engineers need to consider the requirement under analysis and the model behaviors, and guide the search toward portions of the input domains that more likely contain failure-revealing test cases. 
We designed manual fitness functions by reverse engineering the models, considering each requirement, and identifying portions of the input domains that we believed more likely contained failure-revealing test cases. 
\NAMEMANUAL is executed for each assumption-requirement combination — when a specific  combination is considered, the automatically-generated and manually-derived fitness functions associated with that requirement are considered. 
Therefore, the manual fitness function designed for one requirement cannot harm the search with respect to another requirement.
Even if manual fitness functions are defined for each requirement, manual fitness functions can be reused (and be useful) for multiple requirements.
For example, as shown in \Cref{tab:inputs}, the requirements CC1 and CC3 share the same manual fitness function.

\textbf{Applicability of the automatic fitness functions}: Although \NAMESIMULINK considers automatic fitness functions obtained from requirements specified in STL, our approach is largely applicable.
\simulink provides several tools that enable engineers to formalize requirements, such as Test Assessment blocks~\cite{TestAssessment}, which were also considered as inputs for SBST of CPS~\cite{formica2022simulation}.
\NAME can be extended to support any other automatic fitness function.
For example, we plan to integrate \NAMESIMULINK with an approach we recently proposed that automatically converts Test Assessment blocks into automatic fitness functions~\cite{formica2022simulation}.
This extension will enable our framework to support requirements specified using Test Assessment blocks that are natively integrated into \simulink.

\NAME can be instantiated using different automated translations from STL to fitness functions. Our implementation, \NAMESIMULINK relies on the translation used by \staliro and inherits the same limitations. As previously mentioned, we selected \staliro due to its recent classification as ready for industrial development, and its use in several industrial systems. Our solution can be extended by considering other automated translations from the literature (e.g., \cite{fainekos2009robustness,fainekos2006robustness,Donze2010,Pant2017,menghi2019generating}). 
An extensive comparison of these solutions is out of scope.

Automatically-generated fitness functions for multiple requirements can be obtained via \NAME by writing an STL formula representing the Conjunction of the requirements and running the automated translation.
For example, the automatically-generated fitness function for requirement AT6abc is obtained by generating the conjunction of requirements AT6a, AT6b, and AT6c and by running the automated translation.
These requirements can benefit from dedicated automated fitness functions, such as \cite{Porres2022}.

\textbf{Applicability of \NAME}: Our running example concerning the automatic transmission model from \Cref{sec:applSimulink} is from the automotive domain. However, as shown in our evaluation (\Cref{sec:evaluation}), \NAME is applicable in many CPS domains, including energy~\cite{jin2014powertrain}  and medical~\cite{sankaranarayanan2012simulating}.
\Cref{sec:evaluation} presents both the automatically-generated and manually-derived fitness functions considered for our benchmark models.
Other instances of \NAME can target different types of artifacts, e.g., source code or other types of models.
Finally, our results show the benefits and complementarity of automatically-generated and manually-derived fitness functions in the context of SBST of CPS:  even if a formal requirement was available for CC4, the search driven by either the automatic or the manual fitness functions did not return any failure-revealing test cases. 
On the contrary, the combination of the two returned a set of failure revealing test cases.

The scalability of the proposed technique depends on many factors: the number of input signals, the number of control points, their input ranges, the number of blocks  of the model and the time required to simulate it.
We can not claim that a higher number of input signals will necessarily negatively impact the scalability of the proposed technique, since those additional input signals may simplify the detection of failure revealing test cases. 
More empirical studies, like the one presented in this paper, can assess how the number of input signals, the number of control points, their input ranges, the number of blocks of the model and the time required to simulate it influence the scalability of the SBST frameworks.

Our results provide empirical evidence supporting the usefulness of manual fitness functions.
This result is not obvious: the use of manual fitness functions can worsen the search process. 
For example, \NAME did not improve the search process compared to \staliro for one of the benchmarks (i.e., NN-R), and, for another, using \NAME worsens the performance compared to \staliro (i.e., WT3-R). 
Therefore, our results are significant. They provide empirical evidence that, for our benchmark, the manual fitness functions improved the search process.

The first author designed the manual fitness functions we considered in our experimentation.
The preliminary experiment to assess the effort required to write manual fitness functions relies on two subjects. 
Although our study considers a limited number of participants, it is one of the first attempts to perform such a user study for SBST tools in the \simulink domain.
Future work can extend our analysis by considering a larger number of participants and investigating the tool's usability by end-users in this domain.

The general framework presented in \Cref{sec:Athena} can be instantiated by considering different fitness functions, including branch distance (see for example~\cite{arcuri2010does}). 
In our implementation (\Cref{sec:implementation}), the fitness function automatically generated from the requirement relies on the notion of robustness.
Similarly to branch distance, robustness evaluates how far a predicate is from obtaining its opposite value.
However, unlike branch distance, the robustness metric is evaluated on predicates that change their value over time, and therefore provide the semantics for logical temporal operators, such as globally and eventually. 

\NAME is a framework for black-box testing: it assumes that the internal structure of the model is unknown. The search algorithm explores the input domain looking for failure-revealing test cases. While fitness landscape analysis (e.g.,~\cite{DBLP:journals/ase/AletiMG17,campos2018empirical,albunian2020causes,joffe2019directing}) can help engineers in designing the manual fitness function, we did not use it not to bias our evaluation; using the fitness landscape analysis to support the design of the manual fitness function is an interesting direction for future work that can further improve our results and is outside the scope of this work.

\textbf{External Validity:} 
The selection of the models and requirements is a threat to the external validity of our results.
However, the benchmark models considered in \emph{RQ1} and \emph{RQ2} 
(a)~were extensively used in the SBST literature (e.g.,~\cite{DBLP:conf/arch/ErnstABCDFFG0KM21,fehnker2004benchmarks,jin2014powertrain,Aristeo,Fainekos2019,Donze2010}), 
(b)~are representative of different CPS systems, i.e., 
AT, AFC and CC are from the automotive domain,
NN is from the ML domain, 
WT is from the electrical domain, 
and F16 is from the aerospace domain,
(c)~some of the models were developed by engineers working in the industry, i.e., AFC is from Toyota~\cite{jin2014powertrain}.
The case studies used to answer \emph{RQ5} are large and complex case studies representative of industrial systems developed by MathWorks~\cite{MathWorks} and 
linked to the EcoCAR Mobility Challenge~\cite{EcoCAR}, a competition sponsored by the U.S. Department of Energy~\cite{DOE}, General Motors~\cite{GM}.

The reverse-engineering process we used to define the manual fitness functions for \NAMEMANUAL (see Section~\ref{sec:rq1}), the fitness function of \NAMESIMULINK, and the fault we injected are a threat to the external validity of our results. 
However, the results of \NAMEMANUAL and \NAMESIMULINK are likely to improve when the designers of the fitness functions are knowledgeable about the domain and engineered the \simulink models.

\NAMESIMULINK explicitly targets \simulink models. 
Other instances of \NAME can target different model types or software programs.
We conjecture that our results are also generalizable to these domains. 
More empirical studies, such as those presented in this paper, will determine whether our conjecture holds. 

The selection of the automatically-generated, manually-derived, and \NAME fitness functions threat the external validity of our results.
Although the \NAME fitness function relies on the value of the parameter $p$ to linearly combine the value of automatically-generated and manually-derived fitness functions, other \NAME fitness functions can be used by engineers (as detailed in \Cref{sec:implementation}).
\Cref{sec:athenafitnessinfluence} experiments with different values of $p$ and discusses how we selected the value of $p$.

\textbf{Internal Validity:} 
The values assigned to the configuration parameters of our tools are a threat to the internal validity of our results. 
However, our configuration does not favor any of the frameworks as the common configuration parameters of \NAMEAUTOMATIC, \NAMEMANUAL, and \NAME share the same values. 

The current experiments demonstrate that \NAME outperforms \NAMEAUTOMATIC and \NAMEMANUAL. 
While it is possible to address a search problem by manually defining ad-hoc fitness functions in theory, \Cref{sec:manualfitnessinfluence} showed that this was also possible in practice.  
The procedure we used to define the manual fitness functions (\Cref{sec:manualfitnessinfluence}) required us to reverse engineer the model. 
We were not the designers of the models. 
\Cref{sec:athenafitnessinfluence} defined the procedure we used to select the value for the parameter $p$ of the \NAME fitness function.
In practice, engineers are more knowledgeable about their models and can select more appropriate values for the parameter $p$ depending on many factors, such as the development stage and the likelihood of finding failure by prioritizing the manual and the automatic fitness functions.
During earlier development stages, engineers will likely prioritize explorative search by relying on the automatic fitness function and move toward exploitative search that depends on the manual fitness function as the design progresses and they become more aware of areas of the input domain that are more critical for the satisfaction of the requirement.
Therefore, the methodology we followed in our evaluation penalizes our approach, and the current experimental method confirms the hypothesis on the effectiveness of \NAME.


\section{Related Work}
\label{sec:RelWorks}
Despite the vast research literature on SBST and the considerable number of surveys on the topic (e.g.,
~\cite{5210118,DBLP:journals/soco/KhariK19,NELSON2009345,DBLP:conf/pts/KluckZWN19,7102580,brunetto2021introducing}), we did not find any existing work classifying fitness functions into automatically generated and manually defined. 
In addition, we are not aware of any work proposing a framework that combines these two types of fitness functions. 
Therefore, this section summarizes related work targeting \emph{automatically generated} fitness functions, \emph{manually defined} fitness functions, and \emph{multi-objective} fitness functions.

\emph{Automated generation} of fitness function
techniques often compute fitness functions from logic-based specifications (e.g.,~\cite{fainekos2009robustness,fainekos2006robustness,Donze2010,Pant2017,menghi2019generating}). 
Well established SBST tools, such as \Breach~\cite{Breach}, \staliro~\cite{S-Taliro}, \Aristeo~\cite{Aristeo}, and \FalStar~\cite{Falstar},  rely on these fitness functions.
For each atomic proposition, these fitness functions typically compute a value indicating a satisfaction degree for the proposition at every time instant.
There are alternative ways to compute fitness values associated with temporal operators, such as the use of $\texttt{min}$ and $\texttt{max}$ operators~\cite{fainekos2009robustness,fainekos2006robustness},
distance operators computing smooth approximations of the $\texttt{min}$ and $\texttt{max}$ operators (e.g.,~\cite{Pant2017}), arithmetic and geometric mean along a time interval (e.g.,~\cite{Mehdipour2019,Lindemann2019,Varnai2020}), cumulative values over a time horizon (e.g.,~\cite{Haghighi2019}). 
Some approaches also consider perturbations that may shift the signal values over time to compute the fitness value (e.g.,~\cite{Donze2010}).

A variety of \emph{manually defined} fitness functions was proposed in the literature.
Manually defined fitness functions can guide the search to produce test outputs with diverse shapes~\cite{7886937}, maximize or minimize the outputs of a system characterizing its critical behavior~\cite{8453180}, maximize diversity in output signals~\cite{matinnejad2016automated}, quantify the difference between a reference and an output signal~\cite{arrieta2021using}, and minimize the difference between the expected and simulated behavior of a CPS~\cite{9476233}.
Many manually defined fitness functions were also proposed to guide SBST frameworks that analyze software code. 
For example, some recent manually defined fitness functions measure line coverage, input coverage, output coverage  (e.g.,~\cite{10.1145/3293882.3330552}),
branch distances (e.g.,~\cite{7840029}), 
test length, method sequence diversity, and crash distance~\cite{9285999},
the cumulative number of defects and the total amount of code to inspect~\cite{10.1145/2908812.2908938}.
Manually defined fitness functions can also compute exploration measures~\cite{10.1145/2642937.2642978}, and 
combine coverage-based and feature interaction measures~\cite{abdessalem2018testing}.

Differently from these works, we proposed a framework that combines automatically generated and manually defined fitness functions and showed the benefits of our framework by considering a large benchmark made by seven models and $27$ requirements and two complex case studies from the automotive and medical domains. 

\emph{Multi-objective} fitness functions are extensively studied by the research literature (see for example ~\cite{li2019quality,wu_2022,sayyad2013pareto,ali2020quality,ramirez2019survey,li2020evaluate,ramirez2018systematic,nuh2021performance} for recent surveys).
For example, SBST frameworks guided by multi-objective search are used for machine learning and automotive systems (e.g.,~\cite{gambi2019automatically,klischat2019generating,haq2022efficient,ben2016testing,riccio2020model}), test case prioritization (e.g.,~\cite{pradhan2019employing,pradhan2018remap}), software refactoring (e.g.,~\cite{mohan2019using}), software energy consumption (e.g.,~\cite{alvarez2016multi}), generation of software assumptions (e.g.,~\cite{9507379,10.1145/3368089.3409737}), and many others uses.
These frameworks use fitness functions based on more than one objective to guide the search exploration.
Within multi-objective fitness functions, optimizing one objective may harm another. 
Therefore, there is no single optimal solution for these problems, but a  set of  non-dominated optimal solutions called Pareto Front~\cite{wright1999numerical}.
Therefore, SBST usually aims to detect (a solution belonging to) the Pareto Front.
Quality Indicators~\cite{li2019quality} are often used to compare the performance of these algorithms and  are chosen depending on the specific application.
SBST frameworks that rely on multi-objective fitness functions also exist for Matlab and \simulink (e.g.,~\cite{TOFFOLO2002}).

Our solution shares some similarities with these works: The fitness function obtained by combining the automatic and the manual fitness functions is a multi-objective fitness function.
However, our work is significantly different since it is the first work that enables the combination of manual and automatic fitness functions.
Proposing a framework that enables engineers to use manual and automatic fitness functions is a significant contribution.
As we have shown in our evaluation, our solution is reusable and applicable in different domains. Such support helps engineers find failure-revealing test cases they can not find with other existing approaches.

\section{Conclusion and Future Work}
\label{sec:conclusion}
We presented \NAME, a novel SBST framework that combines automatically generated and manually defined fitness functions. 
We defined \NAMESIMULINK, an instance of \NAME that supports \simulink models.
We considered two versions of \NAMESIMULINK to assess its benefits in generating failure-revealing test cases.
The two versions of \NAMESIMULINK we considered are preferable over the baseline tools for respectively $\approx79\%$ and $100\%$ of our assumption-requirement combinations.
They generated more failure-revealing test cases than one baseline SBST framework that relies on automatic fitness functions  (+\RQoneathenavsstaliro and 
+\RQonewinsathenabestvsstaliro) and another based on manually defined fitness functions (+\RQoneathenavsmanual and +\RQonewinsathenabestvsmanual). 
\NAMESIMULINK did not show statistically significant differences in efficiency compared with the baseline tools.
Finally, \NAMESIMULINK generated a failure-revealing test case for two large representative automotive and medical case studies.
We discussed the impacts of our results on software engineering practices, and suggested a workflow that combines \NAME and SBST frameworks driven by automatically generated and manually defined fitness functions.

We plan to further support engineers with techniques that automatically infer the value assigned to the parameter $p$ and dynamically change the prioritization of the automatically generated and manually defined fitness functions at run-time in future work.


\section*{Data Availability}
A replication package containing all of our data, test results, and models is publicly available~\cite{Material}.

\begin{acks}
We thank Nicholas Petrunti (McMaster University) for his comments, feedback, and help with the evaluation section of this work. 
We also thank our anonymous study subjects for volunteering to help with RQ1.

We acknowledge the support of the Natural Sciences and Engineering Research Council of Canada (NSERC), [funding reference number RGPIN-2022-04622,DGECR-2022-0040].

This research was enabled in part by support provided by Compute Ontario (\url{www.computeontario.ca}) and Compute Canada (\url{www.computecanada.ca}).
\end{acks}

\bibliographystyle{ACM-Reference-Format}
\bibliography{bibliography}

\end{document}